# Dark Matter and Galaxy Formation

Joel R. Primack[a]

[a]*Physics Department, University of California, Santa Cruz, CA 95064 USA*

**Abstract.** The four lectures that I gave in the XIII Ciclo de Cursos Especiais at the National Observatory of Brazil in Rio in October 2008 were (1) a brief history of dark matter and structure formation in a ΛCDM universe; (2) challenges to ΛCDM on small scales: satellites, cusps, and disks; (3) data on galaxy evolution and clustering compared with simulations; and (4) semi-analytic models. These lectures, themselves summaries of much work by many people, are summarized here briefly. The slides [1] contain much more information.

**Keywords:** Cold dark matter, Cosmology, Dark matter, Galaxies, Warm dark matter
**PACS:** 95.35.+d 98.52.-b 98.52.Wz 98.56.-w 98.80.-k

## SUMMARY

(1) Although the first evidence for dark matter was discovered in the 1930s, it was not until the early 1980s that astronomers became convinced that most of the mass holding galaxies and clusters of galaxies together is invisible. For two decades, theories were proposed and challenged, but it wasn't until the beginning of the 21st century that the ΛCDM "Double Dark" standard cosmological model was accepted: cold dark matter – non-atomic matter different from that which makes up the stars, planets, and us – plus dark energy together making up 95% of the cosmic density. Alternatives such as MOND are ruled out. The challenge now is to understand the underlying physics of the particles that make up dark matter and the nature of dark energy.

(2) The ΛCDM cosmology is the basis of the modern cosmological Standard Model for the formation of galaxies, clusters, and larger scale structures in the universe. Predictions of the ΛCDM model regarding the distribution of galaxies both nearby and out to high redshifts have been repeatedly confirmed by observations. However, on sub-galactic scales there are several potential problems, summarized here under the rubrics satellites, cusps, and angular momentum. Although much work remains before any of these issues can be regarded as resolved, recent progress suggests that all of them may be less serious than once believed.

(3) The goals of cosmology now are to discover the nature of the dark energy and dark matter, and to understand the formation of galaxies and clusters within the cosmic web gravitational backbone formed by the dark matter in our expanding universe with its increasing fraction of dark energy. This third lecture discusses the data on galaxy evolution and clustering both nearby and at high redshifts compared to simulations.

(4) Semi-Analytic Models (SAMs) are still the best way to understand the formation of galaxies and clusters within the cosmic web dark matter gravitational skeleton, because they allow comparison of variant models of star and supermassive black hole formation and feedback. This lecture discusses the current state of the art in semi-analytic models, and describes the successes and challenges for the best current ΛCDM models of the roles of baryonic physics and supermassive black holes in the formation of galaxies.

This paper is a short summary of the lectures. The slides [1] contain much more information.

## Dark Matter Is Our Friend

Dark matter preserved the primordial fluctuations in cosmological density on galaxy scales that were wiped out in baryonic matter by momentum transport (viscosity) as radiation decoupled from baryons in the first few hundred thousand years after the big bang. The growth of dark matter halos started early enough to result in the formation of galaxies that we see even at high redshifts $z > 6$. Dark matter halos provide most of the gravitation within which stable structures formed in the universe. In more recent epochs, dark matter halos preserve these galaxies, groups, and clusters as the dark energy tears apart unbound structures and expands the space between bound structures such as the Local Group of galaxies. Thus we owe our existence and future to dark matter.

Cold dark matter theory [1] including cosmic inflation has become the basis for the standard modern ΛCDM cosmology, which is favored by analysis of the available cosmic microwave background data and large scale structure data over even more complicated variant theories having additional parameters [2]. Most of the cosmological density is nonbaryonic dark matter (about 23%) and dark energy (about 72%), with baryonic matter making up only about 4.6% and the visible baryons only about 0.5% of the cosmic density. The fact that dark energy and dark matter are dominant suggests a popular name for the modern standard cosmology: the "double dark" theory, as Nancy Abrams and I proposed in our recent book about modern cos-mology and its broader implications [5].

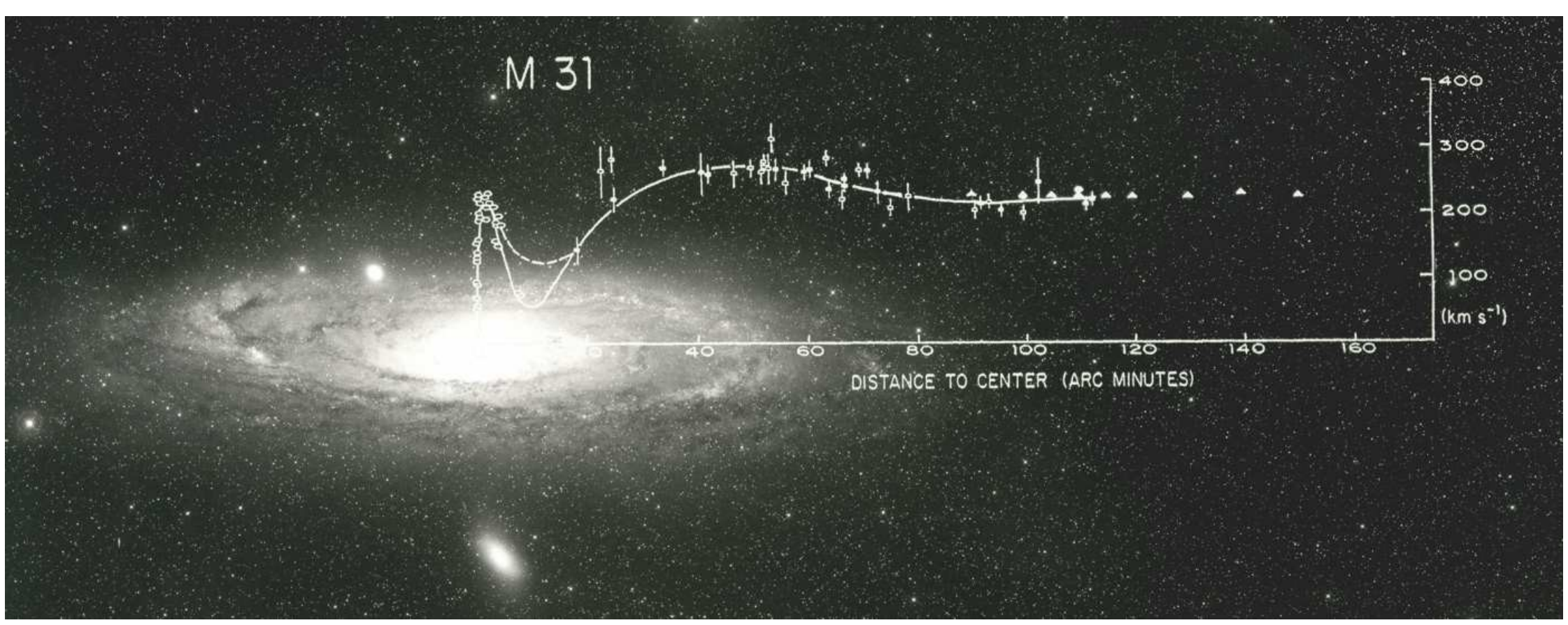

**FIGURE 1.** Optical (dots) and radio (triangles) rotation curve data for the Andromeda galaxy M31. superimposed on the M31 image from the Palomar Sky Survey (from Vera Rubin [3]; see also[4]).

## 1. A BRIEF HISTORY OF DARK MATTER

Table 1 summarizes what people knew about dark matter and when they knew it.

**TABLE 1.** A Brief Chronology of Dark Matter

| | |
|---|---|
| 1930s | Discovery that cluster velocity dispersion ~ 1000 km/s |
| 1970s | Discovery of flat galaxy rotation curves |
| 1980 | Astronomers convinced dark matter binds galaxies and clusters |
| 1980-83 | Short life of Hot Dark Matter theory |
| 1982-84 | Cold Dark Matter (CDM) theory proposed |
| 1992 | COBE discovers CMB fluctuations as predicted by ΛCDM |
| | CHDM and ΛCDM are favored CDM variants |
| 1998 | SN Ia and other evidence of Dark Energy |
| 2000 | ΛCDM is the Standard Cosmological Model |
| 2003- | WMAP and LSS data confirm ΛCDM predictions |
| ~2010 | Discovery of dark matter particles?? |

This is not the place for a detailed historical account with complete references, so instead the early history of dark matter is summarized in Table 2. My lecture slides [1] included key excerpts from many of the early and later dark matter papers, along with photos of their authors. The first lecture ended with an illustrated video version of David Weinberg's "dark matter rap" [6].

**TABLE 2.** Early Papers on Dark Matter

| | |
|---|---|
| 1922 | Kapteyn: "dark matter" in Milky Way disk [7] |
| 1933, 37 | Zwicky: "dunkle (kalte) materie" in Coma cluster |
| 1937 | Smith: "great mass of internebular material" in Virgo cluster |
| 1937 | Holmberg: galaxy mass $5\text{x}10^{11}$ $M_\odot$ from handful of pairs [7] |
| 1939 | Babcock observes rising rotation curve for M31 [7] |
| 1940s | large cluster velocity dispersion $\sigma_V$ confirmed by many observers |
| 1957 | van de Hulst: high HI rotation curve for M31 |
| 1959 | Kahn & Woltjer: MWy-M31 infall $\Rightarrow M_{LocalGroup} = 1.8\text{x}10^{12}\ M_\odot$ |
| 1970 | Rubin & Ford: M31 flat optical rotation curve – see Fig. 1 |
| 1973 | Ostriker & Peebles: halos stabilize galactic disks |
| 1974 | Einasto, Kaasik, & Saar; Ostriker, Peebles, & Yahil summarize evidence for cluster DM & galaxy M/L increase with radius |
| 1975; 78 | Roberts; Bosma: extended flat HI rotation curves |
| 1978 | Mathews: X-rays reveal dark matter of Virgo cluster |
| 1979 | Faber & Gallagher: convincing evidence for dark matter [8] |

The identity of the dark matter remains a key question. The idea that it is low-mass neutrinos was proposed in 1973 by Marx & Szalay and by Cowsik & McClelland, and the theory of structure formation with neutrino dark matter was worked out by Jacob Zel'dovich and his group in the early1980s [9]. Zel'dovich had assumed that the early universe was nearly homogeneous, with a scale-free spectrum of adiabatic fluctuations. By 1980, the upper limit of $(\Delta T/T)_{CMB} < 10^{-4}$ on the fluctuations in the cosmic background radiation temperature in different directions had ruled out the possibility that the matter in the universe is baryonic (i.e., made of atoms and their constituents), and an experiment in Moscow appeared to show that the electron neutrino has a mass of 10s of eV. But this was not confirmed by other experiments, and in 1983 a simulation by White, Frenk, and Davis [10] ruled out light neutrino dark matter by showing that the distribution of galaxies in such a "Hot Dark Matter" (HDM) universe would be much more inhomogeneous than observed. This is because the light neutrinos would remain relativistic until the mass enclosed by the horizon was at least as large as galaxy cluster masses, which would damp smaller scale fluctuations [11].

**TABLE 3.** Early Papers Relevant to Cold Dark Matter

| | |
|---|---|
| 1967 | Lynden-Bell: violent relaxation (also Shu 1978) |
| 1976 | Binney; Rees & Ostriker; Silk: Cooling curves |
| 1977 | White & Rees: galaxy formation in massive halos |
| 1980 | Fall & Efstathiou: galactic disk formation in massive halos |
| 1982 | Guth & Pi; Hawking; Starobinski: Cosmic Inflation $P(k) \propto k^1$ |
| 1982 | Pagels & Primack: lightest SUSY particle stable by R-parity: gravitino |
| 1982 | Blumenthal, Pagels, & Primack; Bond, Szalay, & Turner: WDM |
| 1982 | Peebles: CDM P(k) - simplified treatment (no light neutrinos) |
| 1983 | Milgrom: modified Newtonian dynamics (MOND) alternative to DM |
| 1983 | Goldberg: photino as SUSY CDM particle |
| 1983 | Preskill, Wise, & Wilczek; Abbott & Sikivie; Dine & Fischler: Axion CDM |
| 1983 | Blumenthal & Primack; Bond & Szalay: CDM; WDM P(k) |
| 1984 | Blumenthal, Faber, Primack, & Rees: CDM compared to CfA survey |
| 1984 | Peebles; Turner, Steigman, & Krauss: effects of Λ |
| 1984 | Ellis, Hagelin, Nanopoulos, Olive, & Srednicki: neutralino CDM |
| 1985 | Davis, Efstathiou, Frenk, & White: 1st CDM, ΛCDM simulations |

Pagels and I had suggested in 1982 [12] that the dark matter might be the lightest supersymmetric partner particle, which would be stable because it would be the lightest *R*-negative particle. This particle was likely to be the gravitino (spin 3/2 superpartner of the graviton) in those early days of supersymmetry theory. We showed that the upper limit on the gravitino mass in this case was about 1 keV, and we worked out with Blumenthal implications of this "Warm Dark Matter" scenario for galaxy formation [13]. Steve Weinberg responded to [14] by showing that if a massive gravitino was not the lightest superpartner and was therefore unstable, it could cause serious trouble with big bang nucleosynthesis [15]. This can be avoided if the reheat temperature after cosmic inflation ends is sufficiently low to prevent gravitino formation. Jim Peebles responded to [13] by considering the possibility that the dark matter might be massive enough that it is nonrelativistic in the early universe on all scales relevant for galaxy formation [15]; this would be Cold Dark Matter (CDM).

The "Hot-Warm-Cold" dark matter terminology was introduced by Dick Bond and me in our talks at the 1983 Moriond conference, where I presented early work by George Blumenthal and me on CDM [16]. Key ideas were summarized in Figures 2 and 3. Figure 2 shows that fluctuations of mass less than about $10^{15}$ $M_\odot$ enter the horizon when it is still radiation dominated (i.e., when the scale factor $a < a_{eq}$), and as a result they grow only logarithmically until the universe becomes matter dominated. Fluctuations of greater mass enter the horizon when the universe is matter dominated and as a result they grow as fast as possible, proportional to the scale factor $a$. Consequently, there is a bend in the power spectrum of density fluctuations *P(k)* on length scales corresponding to the comoving horizon size when the universe becomes matter dominated. We showed that a primordial $P(k) \propto k$ power spectrum becomes $\propto k^{-3} (\ln k)^2$ for large $k$ (i.e., small length scales) when the growth of cold dark matter fluctuations is taken into account [17].

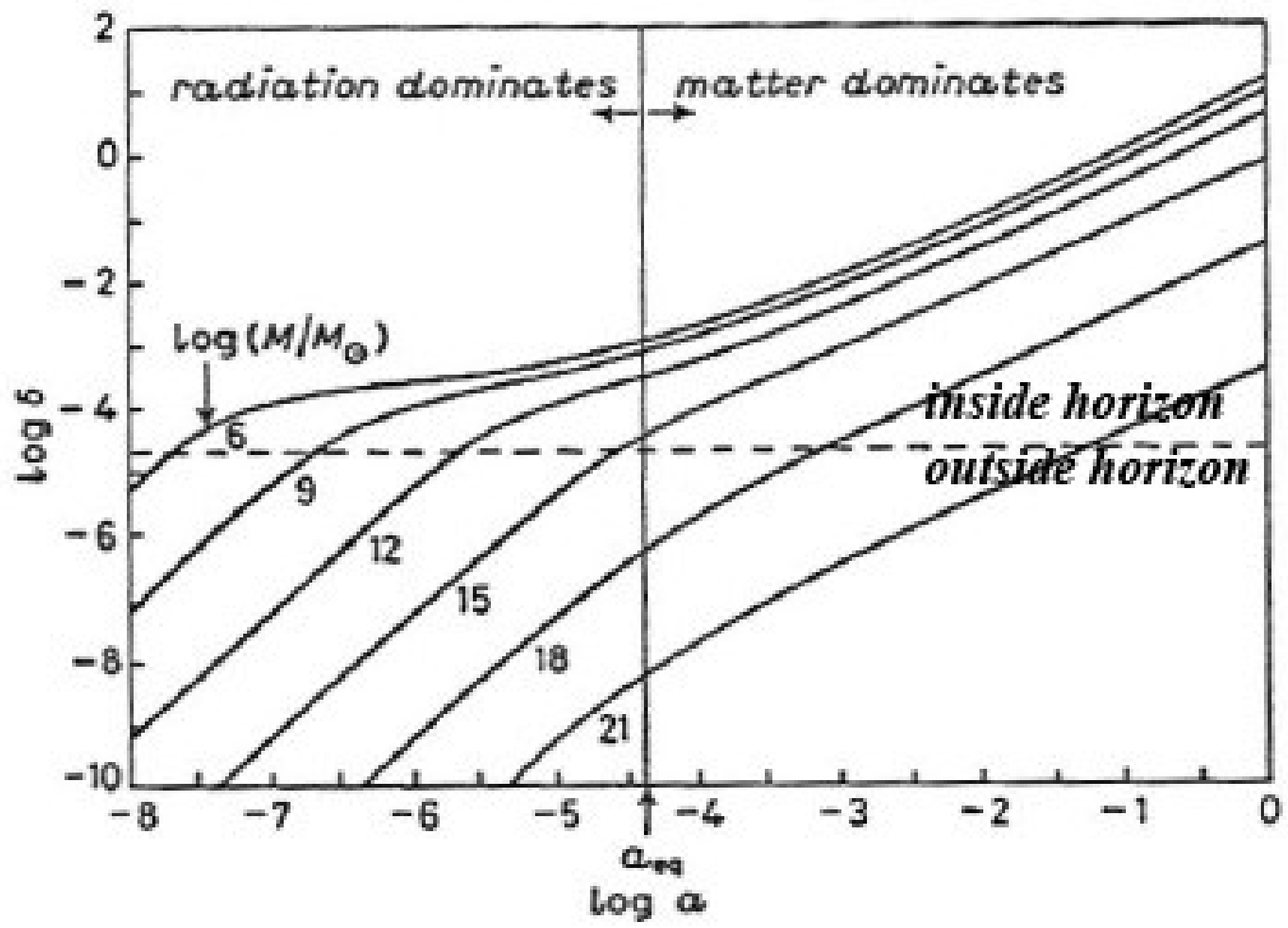

**FIGURE 2.** The growth of the amplitude $\delta = \delta\rho/\rho$ of fluctuations of mass $10^6$, $10^9$, $10^{12}$, $10^{15}$, $10^{18}$, and $10^{21}$ $M_\odot$ in an $\Omega_m = 1$ CDM universe vs. scale factor $a = (1+z)^{-1}$. (From [16].)

Figure 3 shows that fluctuations of total mass (including dark matter) between about $10^8$ and $10^{12}$ $M_\odot$ lie under the cooling curves, i.e. that the cooling time will be shorter than the dynamical time, so that their gravitational collapse will not be impeded by cooling. Thus galaxies should have masses in this range since these CDM fluctuations lie below the cooling curves, while fluctuations of group and cluster masses lie above them. We were enormously encouraged that this implied that CDM could potentially explain the observed mass range of galaxies. The figure furthermore suggested that the range from late to early type galaxies might represent a combination of increasing halo mass and increasing amounts of baryonic dissipation (dashed curve in the figure, with the vertical part representing dissipation within dark matter halos and the bend representing dissipation within the baryon-dominated halo centers. This

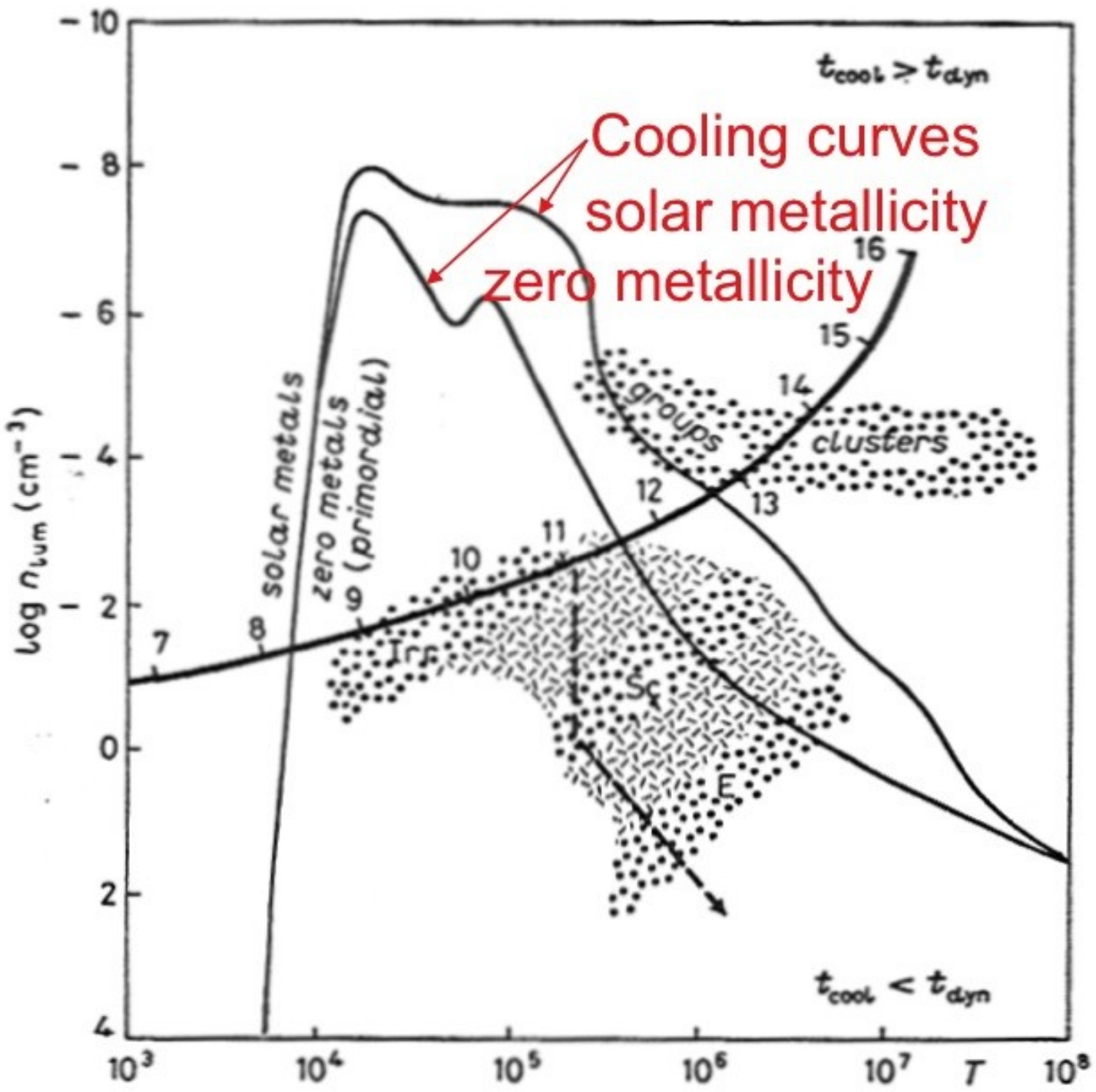

**FIGURE 3.** The baryonic density vs. temperature $T$ as perturbations having total mass M become nonlinear and virialize. Fluctuations below the cooling curves, for primordial and solar metallicity, collapse rapidly since their cooling time $t_{cool}$ is less than their dynamical time $t_{dyn}$. The heavy curve represents typical (1σ) dark matter halos of various masses M: the numbers on the tick marks are $\log_{10}$ (M/$M_{\odot}$). The figure assumes $\Omega_m = h = 1$ and a baryonic-to-total mass ratio of 0.07. (From [13].)

figure, like subsequent semi-analytic models of galaxy formation, made the simplifying assumption that galaxies are the result of spherical gravitational collapse in which the gas and dark matter are heated to the virial temperature.

The first paper that worked out the implications of CDM for the formation of galaxies and clusters was [18]. Two cases were worked out in detail, standard CDM $\Omega_m = 1$ with $h = 0.5$ and open CDM with $\Omega_m = 0.2$ and $h = 1$; CDM with a large cosmological constant (ΛCDM) was discussed but not worked out. On the basis of simple spherical-collapse semi-analytic calculations, these two models were compared to the data on galaxies, groups, and clusters from the first large redshift survey, CfA1, which had just been completed. The paper said "that a straightforward interpretation of the evidence summarized above favors $\Omega = 0.2$, but that $\Omega = 1$ is not implausible." It concluded: "We have shown that a Universe with ~10 times as much cold dark matter as baryonic matter provides a remarkably good fit to the observed Universe. This model predicts roughly the observed mass range of galaxies, the dissipational nature of galaxy collapse, and the observed Faber-Jackson and Tully-Fisher relations. It also gives dissipationless galactic haloes and clusters…. Finally, the cold DM picture seems reasonably consistent with the observed large-scale clustering, including superclusters and voids. In short it seems to be the best model available and merits close scrutiny and testing." I presented an extended summary of the basis and implications of CDM in lectures at the 1984 Enrico Fermi summer school in Varenna, Italy [19].

Peebles [20] and Turner, Steigman, and Krauss [21] worked out some of the consequences of a cosmological constant for the evolution of a CDM universe, and Steigman and Turner [22] coined the clever acronym WIMP (weakly interacting massive particles) for most kinds of hypothetical CDM particles (except axions). Davis, Efstathiou, Frenk, and White [23] ran the first CDM N-body simulations, including standard CDM, open CDM, and ΛCDM. They found that all of these variants could be a good match to the observed distribution of galaxies – see Fig. 4. They found that the peculiar velocities of galaxies are larger than observed for standard CDM with $\Omega = 1$ unless the galaxies are "biased" with respect to the dark matter, i.e. galaxies form only at high peaks of the dark matter density.

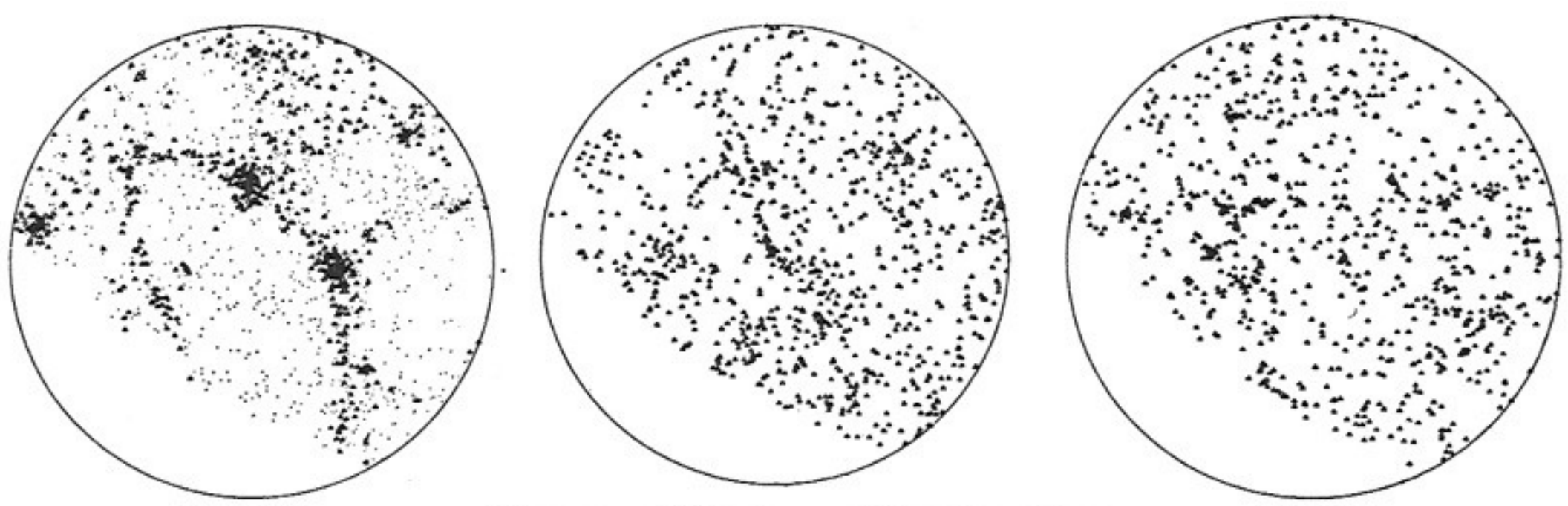

**FIGURE 4.** Early simulations of HDM and CDM compared with the observed galaxy distribution on the sky. The bottom left in each figure represents the part of the sky hidden by galactic obscuration. (From [24].)

However, the discovery in 1986 of large scale flows of galaxies with velocities of order 600 km/s by the "Seven Samurai" team headed by Sandra Faber was inconsistent with significantly biased CDM. But key cosmological parameters – including the Hubble parameter *h*, and the cosmic density parameters for matter ($\Omega_m$) including both cold and hot dark matter, curvature ($\Omega_k$), and vacuum energy ($\Omega_\Lambda$) – were only known very roughly, to within a factor of two or worse. Perhaps all the known cosmological constraints could be satisfied for some set of values of these parameters. Jon Holtzman, in his PhD dissertation research with me, improved the linear fluctuation code that George Blumenthal and I had used and worked out the predictions for cosmic microwave backround anisotropies and other linear effects for 96 CDM variants [25]. By early 1992, we [26,27] had shown that only two CDM variants were consistent with the data then available, namely ΛCDM with $\Omega_m \approx 0.3$ and $\Omega_\Lambda \approx 0.7$, and a mixture of cold and hot dark matter (CHDM) with $\Omega_{cold} \approx 0.7$ and $\Omega_{hot} \approx 0.3$ (so $\Omega_m = 1$ and $\Omega_\Lambda \approx 0$).

At the American Physical Society meeting in April 1992, George Smoot announced the discovery by the Differential Microwave Radiometer (DMR) on NASA's Cosmic Background Explorer (COBE) satellite of fluctuations in the cosmic background radiation temperature in different directions with amplitude $(\Delta T/T)_{CMB} \approx 10^{-5}$. Timothy Ferris quotes me as saying at the time that this ranks as "one of the major discoveries of the century – in fact, it's one of the major discoveries of science" and he quotes Stephen Hawking calling it "the scientific discovery of the century – if not of all time" [27]. We were so enthusiastic because direct evidence of the primordial fluctuations had finally been found, and these fluctuations were consistent with the Harrison-Zel'dovich scale invariant primordial spectrum predicted by cosmic inflation and assumed in CDM models. The amplitude was consistent with the CDM prediction [13] without significant bias. Comparison of the COBE data with Holtzman's predictions [25] for CDM variant models favored the same ΛCDM and CHDM models [29] that we had identified.

Simulations, including those by me and my collaborators [30-33] (cf. [34,35]), showed that ΛCDM and CHDM predicted similar galaxy distributions in the nearby universe. But CHDM, like all models with a critical density of matter (i.e., $\Omega_m = 1$), predicted that galaxies form rather late, while observations increasingly showed the contrary. Also, with the determination by the Hubble Space Telescope Key Project on the Extragalactic Distance Scale that the Hubble parameter $h \approx 0.7$, the time since the Big Bang for $\Omega_m = 1$ was less than 10 Gyr, younger than the oldest stars – which obviously is impossible. Then further discoveries in 1997 clarified the situation. The calibration by the Hipparcos astrometric satellite of the distance scale to the globular clusters in which these stars are found showed that the distance had been underestimated by about 15%, so that the most luminous stars in the oldest globular clusters that are still fusing hydrogen in their cores are 30% brighter, and their age is then about 12±2 Gyr, about 4 Gyr less than had previously been thought. This age of the oldest stars has been independently confirmed by measurement of the depletion by radioactive decay of $^{232}$Th (half life 14.1 Gyr) and $^{238}$U (half life 4.5 Gyr) compared to non-radioactive heavy elements similarly produced by the r-process. These maximum stellar ages are perfectly consistent with the ~ 14 Gyr expansion age of the universe with cosmological densities $\Omega_m = 0.3$ and $\Omega_\Lambda = 0.7$. The essentially simultaneous discovery by the Supernova Cosmology Project and the High-z Supernova Search Team that the expansion of the universe is accelerating clinched the case for $\Omega_m \approx 0.3$ and $\Omega_\Lambda \approx 0.7$.

**TABLE 4.** Some Later Highlights of Cold Dark Matter

| | |
|---|---|
| 1986 | Blumenthal, Faber, Flores, & Primack: baryonic halo contraction |
| 1986 | Large scale galaxy flows of ~600 km/s favor no bias |
| 1989 | Holtzman: CMB and LSS predictions for 96 CDM variants |
| 1992 | COBE DMR discovers $(\Delta T/T)_{CMB} \approx 10^{-5}$, favors ΛCHDM and CHDM |
| 1996 | Seljak & Zaldarriaga: CMBfast code for P(k), CMB fluctuations |
| 1996 | Mo & White: clustering of DM halos |
| 1997 | Navarro, Frenk, & White: DM halo $\rho_{NFW}(r) \propto (r/r_s)^{-1}(1+r/r_s)^{-2}$ |
| 1997 | HST Key Project: $H_0 = 73 \pm 6$ (stat) $\pm 8$ (sys) km/s/Mpc |
| 1997 | Hipparchos distances & SN Ia dark energy $\Rightarrow t_0 \approx 14$ Gyr, $\Omega_\Lambda \approx 0.7$ |
| 2001 | Sheth & Tormen: ellipsoidal collapse gives abundance of DM halos |
| 2001 | Bullock et al.: concentration-mass-redshift relation for DM halos; universal angular momentum structure of DM halos |
| 2002 | Wechsler et al.: halo concentration from mass assembly history |
| 2003- | WMAP and Large Scale Structure surveys confirm ΛCDM predictions |

With calculations focused on this ΛCDM cosmology, the properties of dark matter halos were clarified, in particular, their radial density distribution, clustering, shapes, and evolution with redshift; some of this work was in dissertation research that I supervised by James Bullock, Risa Wechsler, and Brandon Allgood. Increasingly large and carefully controlled galaxy redshift surveys mapped large areas of the nearby universe and small areas of the distant universe, pushing steadily to higher redshifts; this allowed direct measurement of the evolving clustering of galaxies to compare with the ΛCDM predictions. The detailed analyses of the cosmic background radiation temperature and polarization distributions on the sky made possible by the Wilkinson Microwave Anisotropy Probe satellite (WMAP1 [36], WMAP3 [37], and WMAP5 [2,38,39]) and the steadily improving data on the large scale distribution of galaxies from the 2dF and SDSS surveys have confirmed the ΛCDM predictions and determined the cosmological parameters with unprecedented accuracy. The final paragraph of conclusions of a WMAP5 paper [2] says: "Considering a range of extended models, we continue to find that the standard ΛCDM model is consistently preferred by the data. … The CDM model also continues to succeed in fitting a substantial array of other observations." There are now no significant discrepancies between ΛCDM theory and large scale data. The cosmological density parameters are visualized in Fig. 5. The values are now rather precisely known. Maybe someday we will also figure out why they have these values.

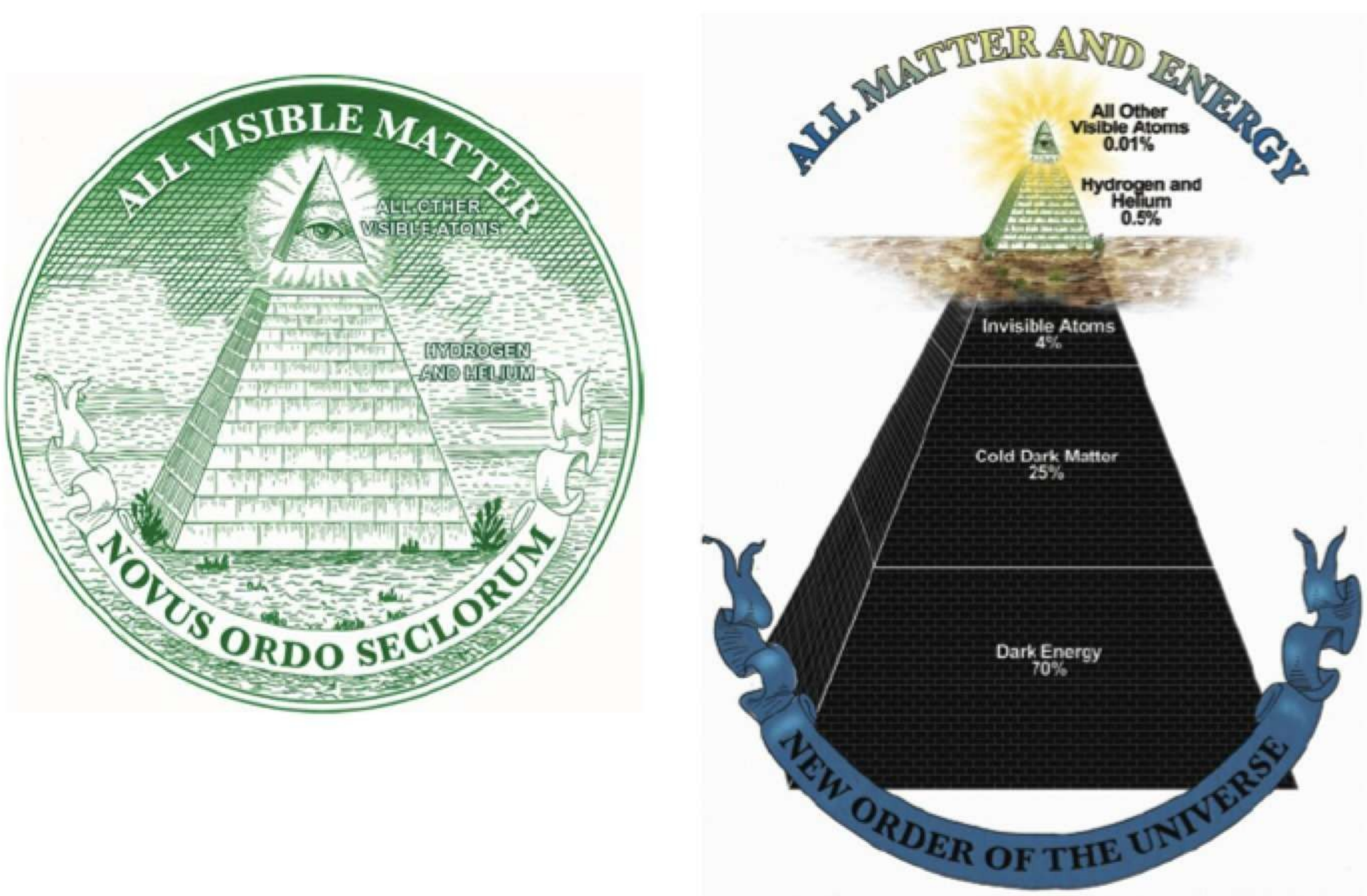

**FIGURE 5**. (left) All visible matter, symbolized by the Great Seal of the United States. Hydrogen and helium make up about 98% of the visible matter, and all heavy elements account for about 2%. (right) The cosmic density pyramid. The visible matter accounts for about 0.5% of the cosmic density, and invisible baryons make up an additional 4%. The latest estimates are that about 23% is cold dark matter and about 72% dark energy. (From [5,40].)

As an example of the multiple cosmological cross-checks now available, there are now five independent paths to determine the cosmic density of baryons: X-ray measurements of galaxy clusters, the relative heights of the first two peaks in the cosmic background radiation angular correlation spectrum, the abundance of deuterium compared to hydrogen in quasar absorption spectra, the absorption of quasar light by the Lyman alpha forest, and the baryon acoustic oscillation (BAO) wiggles in large-scale galaxy correlations. The WMAP5+BAO+SN total cosmic density of baryonic matter is 0.0462±0.0015 [41]. However, the cosmic density of stars and other visible matter is only about 0.005 (with intracluster plasma contributing 0.0018, stars in spheroids and bulges 0.0015, stars in disk and irregular galaxies 0.00055, and stellar remnants 0.0004), according to a review of cosmic matter and energy densities [42].

## What About MOND?

Since the dark matter has not yet been detected except through its gravity and its nature remains a mystery, astrophysicists have naturally considered alternative explanations for the data. No such alternative has yet emerged that is remotely as successful as ΛCDM, but one that has attracted attention is modified Newtonian dynamics (MOND) [43,44]. Besides the fact that MOND is not a predictive cosmological theory and has problems explaining gravitational lensing observations even on galactic scales [45], I want to call attention to four sorts of data that strongly disfavor MOND and support CDM. One is the common observation of galaxies in late stages of merging, in which the dense galactic nuclei have nearly coalesced while the lower-density surrounding material is still found in extended tidal streamers. This is exactly what happens in computer simulations of galaxy mergers, in which the process of dynamical friction causes the massive nuclei to lose kinetic energy to the dark matter and quickly merge. But if there were no dark matter, there would be nothing to take up this kinetic energy and the nuclei would continue to oscillate for a long time, contrary to observation [46]. The second sort of data comes from studies of galaxy clusters [47], including their aspherical shapes that agree well with the predictions of ΛCDM [48,49]. Analysis of X-ray and gravitational lensing data on the "bullet" cluster 1E0657-56 shows particularly clearly that the cluster baryons account for only a small part of the mass, contrary [50,51] to MOND, and also disfavors [52] the "self-interacting dark matter" idea. Gravitational lensing also allows measurement of the mass in clusters on both small and larger scales, and excludes MOND models with neutrino dark matter [53]. The third sort of data is from the relative motion of satellite galaxies about central galaxies, which clearly detects the $\sim r^{-3}$ density decrease at large radii [54]; this is contrary to MOND but predicted by CDM simulations. The fourth sort of data is weak gravitational lensing, which detects mass around red-sequence galaxies as predicted by ΛCDM but not MOND [55]. Thus MOND fails not just on cluster and larger scales but also on galaxy scales.

## Dark Matter Particles

The physical nature of dark matter remains to be discovered. The two most popular ideas concerning the identity of the dark matter particles remain the lightest supersymmetric partner particle [12], also called supersymmetric weakly interacting massive particles (WIMPs) [22], and the cosmological axion [56], recently reviewed in [57]. These are the two dark matter candidate particles that are best motivated in the sense that they are favored by other considerations of elementary particle theory.

Supersymmetry remains the best idea for going beyond the standard model of particle physics. It allows control of vacuum energy and of otherwise unrenormalizable gravitational interactions, and thus may allow gravity to be combined with the electroweak and strong interactions in superstring theory. Supersymmetry also allows for grand unification of the electroweak and strong interactions, and naturally explains how the electroweak scale could be so much smaller than the grand unification or Planck scales (thus solving the "gauge hierarchy problem"). It thus leads to the expectation that the supersymmetric WIMP mass will be in the range of about 100 to about 1000 GeV.

Axions remain the best solution to the CP problem of SU(3) gauge theory of strong interactions, although it is possible that the axion exists and solves the strong CP problem but makes only a negligible contribution to the dark matter density.

Many other particles have been proposed as possible dark matter candidates, even within the context of supersymmetry. An exciting prospect in the next few years is that experimental and astronomical data may point toward specific properties of the dark matter particles, and may even enable us to discover their identity. There are good opportunities for detecting the dark matter particles in deep underground experiments [58], producing them at the Large Hadron Collider, detecting their annihilation products, and exploring the possibility that the dark matter is warm by studying small scale structure (Lecture 2).

## Dark Energy

We can use existing instruments to measure $w = p/\rho$ and see whether it changed in the past. But to get order-of-magnitude better constraints than presently available, anda possible detection of non-cosmological-constant dark energy, better instruments will probably be required both on the ground and in space, according to the Dark Energy Task Force [59]. The National Academy Beyond Einstein report [60] (of which I am a coauthor) recommended the Joint Dark Energy Mission (JDEM) as the first Beyond Einstein mission. It also recommended that JDEM be conceived as a dual-purpose mission, collecting a wide range of data that will shed light on galaxy formation and evolution as well as on the nature dark energy. That way the mission will surely be worth the roughly $1.5 billion that it will cost, even if it turns out not to provide the hoped-for ~3x improvement over future ground-based measurements of dark energy. The amount of improvement depends on the ability to control systematics in new instruments, which is uncertain. At this writing, NASA and DOE are still negotiating their relationship and how to structure the JDEM mission.

# 2. CHALLENGES ON SMALL SCALES: SATELLITES, CUSPS, DISKS

The abundance of dark matter satellites and subhalos, the existence of density cusps at the centers of dark matter halos, and problems producing realistic disk galaxies in simulations are issues that have raised concerns about the viability of the standard cold dark matter (ΛCDM) scenario for galaxy formation. This lecture reviews these issues, and considers the implications for cold vs. various varieties of warm dark matter (WDM). The current evidence appears to be consistent with standard ΛCDM, although improving data may point toward a rather tepid version of ΛWDM – tepid since the dark matter cannot be very warm without violating observational constraints. (This lecture is a substantially updated and expanded version of my talk at the DM08 meeting at Marina Del Rey [61].)

## Subhalos and Satellites

It at first seemed plausible that the observed bright satellite galaxies are hosted by the most massive subhalos of the dark matter halo of the central galaxy, but this turned out to predict too large a radial distribution for the satellite galaxies. Andrey Kravtsov and collaborators [62] proposed instead that bright satellite galaxies are hosted by the subhalos that were the most massive when they were accreted. This hypothesis appears to correctly predict the observed radial distribution of satellite galaxies, and also of galaxies within clusters. It also explains naturally why nearby satellites are dwarf spheroidals (dSph) while more distant ones are a mix of dwarf spheroidal and dwarf irregular galaxies [62].

An issue that is still regularly mentioned by observational astronomers (e.g. [63]) as a problem for ΛCDM is the fact that many fewer satellite galaxies have been detected in the Local Group than the number of subhalos predicted. But developing theory and the recent discovery of many additional satellite galaxies around the Milky Way and the Andromeda galaxy suggest that this is not a problem at all (e.g. [64]). As Fig. 6 (a) shows, it is only below a circular velocity ~30 km $s^{-1}$ that the number of dark matter halos begins to exceed the number of observed satellites. Figure 6 (b) shows that suppression of star formation in small dwarf galaxies after reionization can account for the observed satellite abundance in ΛCDM, as suggested by [65-68]. Whether better understanding of such baryonic physics can also explain the recent discovery [69] that all the local faint satellites have roughly the same dynamical mass of about $10^7$ solar masses within their central 300 parsecs remains to be seen. Alternatively, it is possible that this reflects a clustering scale in the dark matter, which would be a clue to its nature. The newly discovered dwarf satellite galaxy properties such as metallicity appear to continue the scaling relations discovered earlier, with metallicity decreasing with luminosity [70]. Explaining this is another challenge [71-73] for theories of the formation of satellite galaxies.

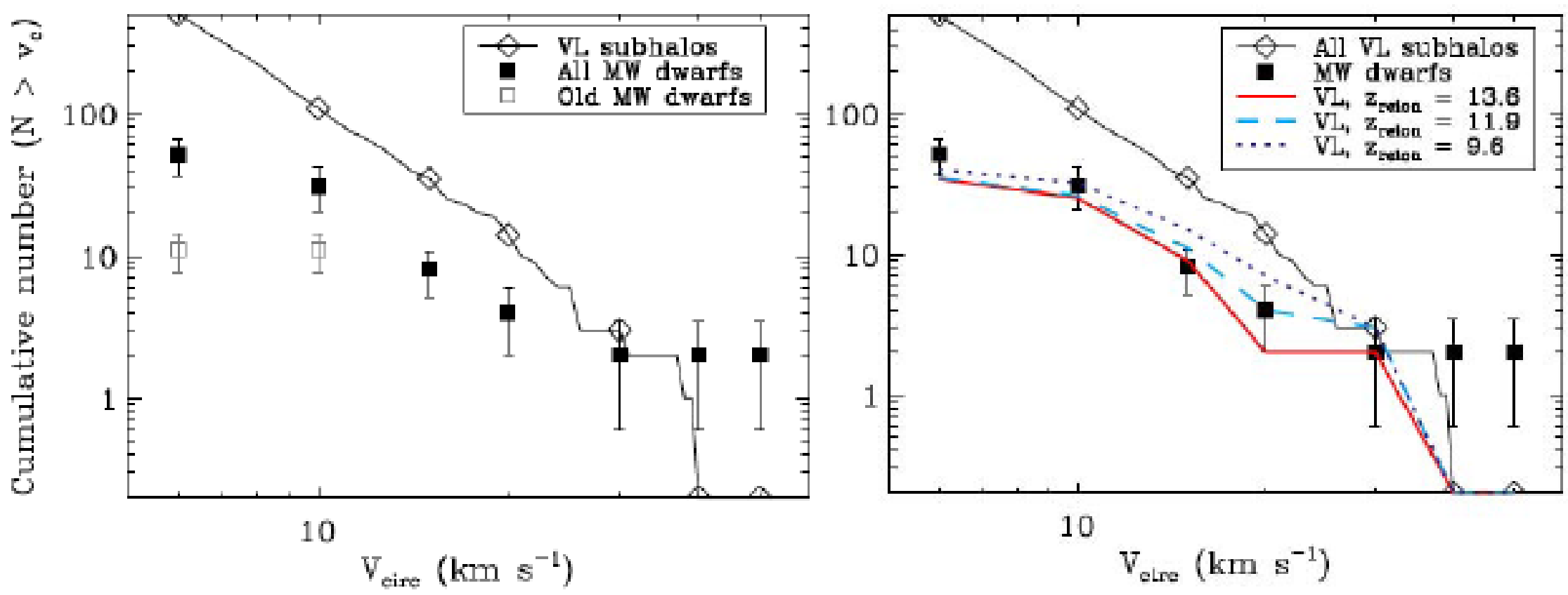

**FIGURE 6.** (a) Cumulative number of Milky Way satellite galaxies as a function of halo circular velocity, assuming Poisson errors on the number count of satellites in each bin. The filled black squares include the new circular velocity estimates from [64], who follow [74] and use $V_{circ} = \sqrt{3}\ \sigma$. Diamonds represent all subhalos within the virial radius in the Via Lactea I simulation [75]. (b) Effect of reionization on the missing satellite problem. The lower solid curve shows the circular velocity distribution for the 51 most massive Via Lactea subhalos if reionization occurred at $z = 13.6$, the dashed curve at $z = 11.9$, and the dotted curve at $z = 9.6$. (Figures from [64].)

Hogan and Dalcanton [76] introduced the parameter $Q = \rho/\sigma^3$ as an estimate of the coarse-grained phase-space density of the dark matter in galaxy halos. Liouville's theorem implies that observed values of Q set a hard lower limit on the original phase-space density of the dark matter. All of the galaxies except UMa I, CVn I, and Hercules have $Q > 10^{-3}\ M_{\odot}\ pc^{-3}\ (km\ s^{-1})^{-3}$, about an order of magnitude improvement compared to the previously-known dSphs. The subhalos in Via Lactea II [77] that could host Milky Way satellites have densities and phase space densities comparable to these values. This places significant limits on non-CDM dark matter models; for example, it implies that the mass of a WDM particle must be $m_x > 1.2$ keV.

The Via Lactea II [77], GHALO [78], and Aquarius simulations [79,80] are the highest resolution simulations of a Milky Way mass halo yet published, and they are able to resolve substructure even at the distance of the sun from the center of the Milky Way. An important question is whether the fraction of mass in the subhalos of mass $\sim 10^6 - 10^8\ M_{\odot}$ is the amount needed to explain the flux anomalies observed in "radio quads" – radio images of quasars that are quadruply gravitationally lensed by foreground elliptical galaxies. A recent paper [80] based on the Aquarius simulations finds that there is probably insufficient substructure unless baryonic effects improve subhalo survivability (see Part 3), and I understand that the Via Lactea group is reaching similar conclusions. Free streaming of WDM particles can considerably dampen the matter power spectrum in this mass range, so a WDM model with an insufficiently massive particle (e.g., a standard sterile neutrino $m_\nu < 10$ keV) fails to reproduce the observed flux anomalies [82]. In order to see whether this is indeed a serious constraint for WDM and a triumph for CDM, we need more than the few radio quads now known – a challenge for radio astronomers! We also need better observations and modeling of these systems to see whether subhalos are indeed needed to account for the flux anomalies in all cases [83-85]. Observing time delays between the images can help resolve such issues [86,87].

An additional constraint on WDM comes from reionization. While the first stars can reionize the universe starting at redshift $z > 20$ in standard ΛCDM [88], the absence of low mass halos in ΛWDM delays reionization [89]. Reionization is delayed significantly in ΛWDM even with WDM mass $m_x = 15$ keV [90]. The actual constraint on $m_x$ from the cosmic microwave background and other data remains to be determined. If the WDM is produced by decay of a higher-mass particle, the velocity distribution and phase space constraints can be different [91,92]. MeV dark matter, motivated by observation of 511 keV emission from the galactic bulge, also can suppress formation of structure with masses up to about $10^7$ $M_\odot$ since such particles are expected to remain in equilibrium with the cosmic neutrino background until relatively late times [93].

Sterile neutrinos that mix with active neutrinos are produced in the early universe and could be the dark matter [94]. Such neutrinos would decay into X-rays plus light neutrinos, so non-observation of X-rays from various sources gives upper limits on the mass of such sterile neutrinos $m_s < 3.5$ keV. Since this upper limit is inconsistent with the lower limit $m_s > 28$ keV from Lyman-alpha-forest data [95], that rules out such sterile neutrinos as the dark matter, although other varieties of sterile neutrinos are still allowed and might explain neutron star kicks [96,97].

Note finally that various authors [98-100] have claimed that ΛWDM substructure develops in simulations on scales below the free-streaming cutoff. If true, this could alleviate the conflict between the many small subhalos needed to give the observed number of Local Group satellite galaxies, taking into account reionization and feedback, and needed to explain gravitational lensing radio flux anomalies. However Wang and White [101] recently showed that such substructure arises from discreteness in the initial particle distribution, and is therefore spurious.

As a result of the new constraints just mentioned, it follows that the hottest varieties of warm dark matter are now ruled out, so if the dark matter is not cold (i.e., with cosmologically negligible constraints from free-streaming, as discussed in the original papers that introduced the hot-warm-cold dark matter terminology [16-18]) then it must at least be rather tepid.

## Cusps in Galaxy Centers

Dark matter cusps were first recognized as a potential problem for CDM by Flores and me [102] and by Moore [103]. However, beam smearing in radio observations of neutral hydrogen in galaxy centers was significantly underestimated [104,105] in the early observational papers; taking this into account, the observations imply an inner density $\rho(r) \propto r^{-\alpha}$ with slope satisfying $0 \leq \alpha < 1.5$, and thus consistent with the ΛCDM Navarro-Frenk-White [106] slope $\alpha$ approaching 1 from above at small radius $r$. The NFW formula $\rho_{NFW}(r) = 4\, \rho_s\, x^{-1} (x + 1)^{-2}$ (where $x = r/r_s$, and the scale radius $r_s$ and the density $\rho_s$ at this radius are NFW parameters) is a rough fit to the dark matter radial density profile of pure dark matter CDM halos. The latest very high resolution simulations of pure dark matter Milky-Way-mass halos give results consistent with a power law central density with α slightly greater than 1 [77] but perhaps with indications of α decreasing at smaller radii [78]. Low surface brightness galaxies are mainly dark matter, so complications of baryonic physics are minimized but could still be important [108,109]. A careful study of the kinematics of five nearby low-mass spiral galaxies found that four of them had significant non-circular motions in their central regions; the only one that did not was consistent with $\alpha \approx 1$ [110] as predicted by ΛCDM for pure dark matter halos. The central non-circular motions observed in this galaxy sample and others could be caused by nonspherical halos [111,112]. Dark matter halos are increasingly aspherical at smaller radii, at higher redshift, and at larger masses [113-116]. This halo asphericity can perhaps account for the observed kinematics [117-120], although analysis of a larger set of galaxies suggests that this would implausibly require nonrandom viewing angles [121].

Recent observations of nearby galaxies combining THINGS HI kinematic data and Spitzer SINGS 3.6 μm data to construct mass models [122] indicate that a core-dominated halo with pseudo-isothermal central profile $\rho(r) \propto (r_0^2 + r^2)^{-1}$ is clearly preferred over a cuspy NFW-type halo for many low-mass disk galaxies, even after correcting for noncircular motions [123]. These and other observations [124] favor a kpc-size core of roughly constant density dark matter at the centers of low-mass disk galaxies.

Only self-consistent ΛCDM simulations of galaxies including all relevant baryonic physics, which can modify the central dark matter density distributions and thus the kinematics, will be able to tell whether ΛCDM galaxies are inconsistent with these observations. Attempts to include relevant baryonic physics have found mechanisms that may be effective in erasing a NFW-type dark matter cusp, or even preventing one from ever forming. At least four such mechanisms have been proposed: (1) rapid removal ("blowout") of a large quantity of central gas due to a starburst causing the dark matter to expand [e.g., 125], and energy and angular momentum transfer to the central dark matter through the action of (2) bars [126], (3) gas motion [e.g. 127], and (4) infalling clumps via dynamical friction [128-130]. Proposal (1) is supported by recent cosmological simulations of formation of small spiral galaxies (F. Governato et al., in preparation). Recent high-resolution simulations [e.g., 131] do not favor (2). But recent work has suggested (3) that supernova-driven gas motions could smooth out dark matter cusps in very small forming galaxies as a consequence of resonant heating of dark matter in the fluctuating potential that results from the bulk gas motions [132], and thus explain observations suggesting dark matter cores in dwarf spheroidal (dSph) galaxies such as the Fornax and Ursa Minor satellites of the Milky Way. These authors suggest that the same mechanism can explain other puzzling features of dSph galaxies, such as the stellar population gradients, the low decay rate for globular cluster orbits, and the low central stellar density. These authors also argue that once the dark matter cusp is smoothed out by baryonic effects in protogalaxies, subsequent merging will not re-create a cusp even in larger galaxies [cf. 133]. Bulk gas motion driven by active galactic nuclei (AGN) has also been shown to be a possible explanation for dark matter and stellar cores in massive stellar spheroids [134].

Recent work also suggests (4) that dynamical friction could explain the origin of dark matter cores in dwarf spheroidal galaxies [135,136] and in low-mass disk galaxies [137,138]. The latter papers compare ΛCDM pure dark matter (PDM) and dark matter + baryons (BDM) simulations starting from the same initial conditions consistent with WMAP3 cosmological parameters. The hydrodynamic BDM simulation includes star formation and feedback. At high redshifts *z* > 7, the PDM and BDM density profiles are very similar. Adiabatic contraction [139-144] subsequently causes the BDM halo to become more cuspy than the PDM one, but then dynamical friction causes infalling baryon+DM clumps to transfer energy and angular momentum to the dark matter. The resulting DM radial profile is essentially pseudo-isothermal with a flat core – see the low-*z* curves in Fig. 7: in the inner ~2 kpc, ρ(R) becomes flat (left panel). The behavior of the dark matter velocity dispersion $\sigma_{DM}$ in the PDM vs. BDM models mirrors that of the density. The NFW cusp in the PDM simulation forms early and is characterized by a "temperature inversion": $\sigma_{DM}(R)$ rising to R~10 kpc. But in the BDM simulation there is no temperature inversion, and indeed $\sigma_{DM}(R)^2 \sim R^{-\beta}$ with β increasing until about *z* ~ 0.6 and decreasing sharply thereafter; this is apparently caused by dynamical friction heating the central DM, causing it to stream outward. The number of subhalos in this inner region of the BDM simulation is about twice that of the PDM simulation, which could be relevant for explaining the anomalous flux ratios in radio quads (discussed in the previous section). The central density distribution in the BDM simulation may be what is needed to explain strong lensing statistics [145]. These very intriguing simulation results need to be confirmed and extended by higher resolution simulations of many more galaxies.

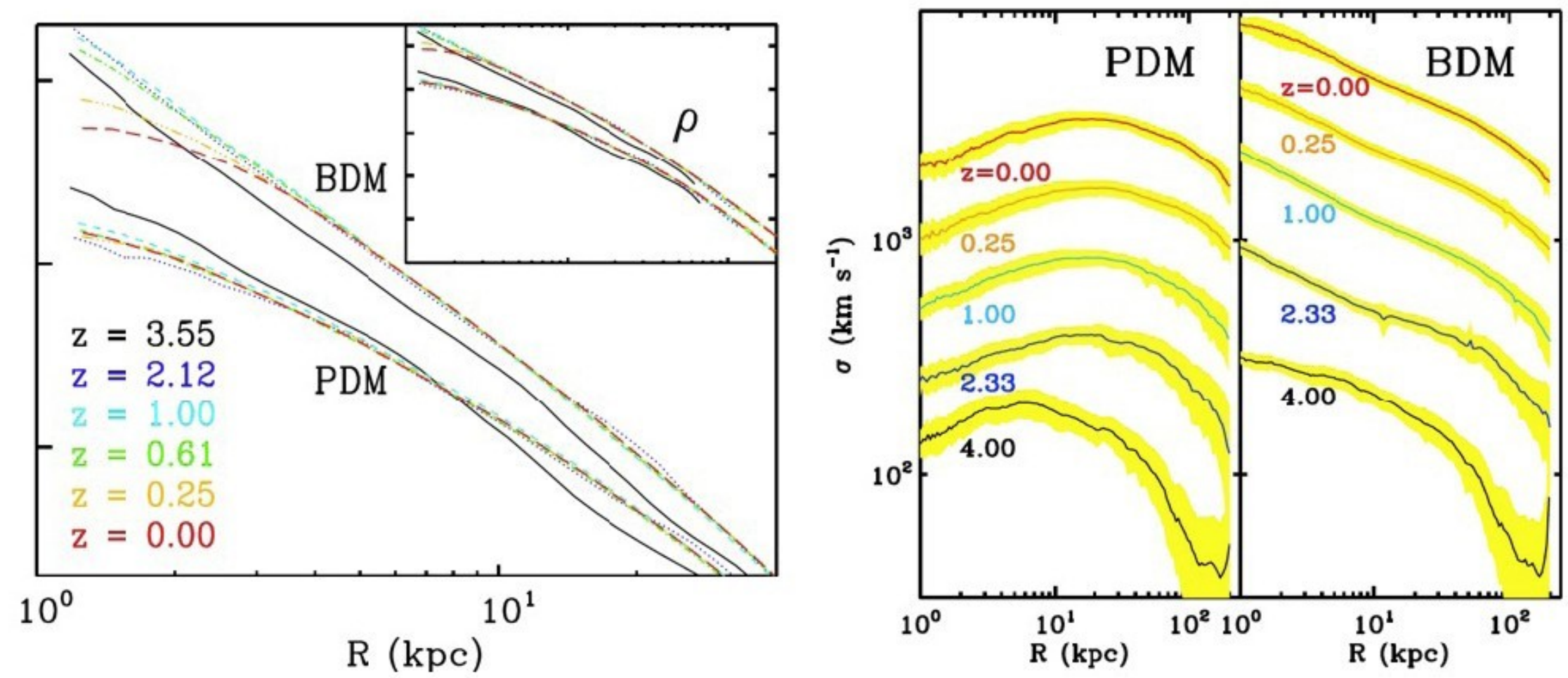

**FIGURE 7. (left)** Redshift evolution of DM density profiles ρ(R) in PDM and BDM models: z = 3.55 (solid), 2.12 (dotted), 1.0 (dashed), 0.61 (dot-dashed), 025 (dot-dash-dotted) and 0 (long dashed). The PDM and BDM curves are displaced vertically for clarity. The inner 40 kpc of halos are shown. The vertical coordinate units are logarithmic and arbitrary. For the PDM model, the density is well fitted by the NFW profile over a large range in *z*, and $r_s$~28 kpc at *z* = 0. For the BDM model, the NFW fit is worse and $R_{iso}$~15 kpc at the end. The insert provides ρ within 200 kpc for comparison. **(right)** Redshift evolution of DM velocity dispersions in PDM and BDM models. Except for the lowest ones, the curves are displaced vertically up for clarity. The second curves from the bottom are displaced by a factor of 2, the third — by a factor of $2^2$, the fourth — by a factor of $2^3$, and the last ones— by a factor of $2^4$. The colored width represents a 1σ dispersion around the mean. The inner 200 kpc of halos are shown. The vertical coordinate units are logarithmic. (From [137].)

Observations indicated that dark matter halos may also be too concentrated farther from their centers [146] compared to ΛCDM predictions. Halos hosting low surface brightness galaxies may have higher spin and lower concentration than average [147,116], which would improve agreement between ΛCDM predictions and observations. As we have just discussed, it remains unclear how much adiabatic contraction [139-144] occurs as the baryons cool and condense toward the center, since there are potentially offsetting effects from gas motions [127] and dynamical friction [137]. Recent analyses comparing spiral galaxy data to theory conclude that there is little room for adiabatic contraction [148,149], and that a bit of halo expansion may better fit the data [149]. Early ΛCDM simulations with high values $\sigma_8 \sim 1$ of the linear mass fluctuation amplitude in spheres of 8 $h^{-1}$ Mpc (a measure of the amplitude of the power spectrum of density fluctuations) predicted high concentrations [150], which are lower with lower values of $\sigma_8$ [151]. The cosmological parameters from WMAP5 and large scale structure observations [2,38,41,152], in particular $\sigma_8 \approx 0.82$, lead to concentrations that match galaxy observations better [153], and they may also match observed cluster concentrations [154,155].

## Galactic Disks

The growth of the mass of dark matter halos and its relation to the structure of the halos has been studied based on structural merger trees [147], and the angular momentum of dark matter halos is now understood to arise largely from the orbital angular momentum of merging progenitor halos [156,157]. But it is now clear that the dark matter and baryonic matter in disk galaxies have very different angular momentum distributions [158,159]. Although until recently simulations were not able to account for the formation and structure of disk galaxies, simulations with higher resolution and improved treatment of stellar feedback from supernovae are starting to produce disk galaxies that resemble those that nature produces [160,161], with rotation velocity consistent with the Tully-Fisher relation between rotation velocity and luminosity or baryonic mass. High-resolution hydrodynamical simulations also appear to produce thick, clumpy rotating disk galaxies at redshifts $z > 2$ [162], as observed [163,164]. It remains to be understood how the gas that forms stars acquires the needed angular momentum. Possibly important is the recent realization that a significant amount of gas enters lower-mass halos cold and in clouds or streams [165-167], rather than being heated to the halo virial temperature as in the standard treatment used in semi-analytic models [18,168].

Once thin stellar disks form, they are in danger of being thickened by mergers. One expects major mergers to be more common for larger mass galaxies because the increasing inefficiency of star formation in higher mass halos limits the total stellar masses of galaxies [169]. Studies of mergers in simulations show that for Milky Way mass galaxies, the largest contribution in mass comes from mergers with a mass ratio of ~1:10 [167]. Thin disks are significantly thickened by such mergers [170], although if the merging galaxies are gas rich, a relatively thin disk can re-form [171-174]. That the majority of large mergers onto $\sim 10^{12}$ $M_{\odot}$ halos are gas rich while the gas fraction decreases for more massive halos $>10^{12.5}$ $M_{\odot}$ [175] could help to explain the increasing fraction of large stellar spheroids in larger mass halos [176]. In the absence of good statistics on the disk thickness of galaxies and the relative abundance of bulgeless disks as a function of galaxy mass, the Sérsic index is a useful proxy. For Milky Way mass galaxies ($V_{rot} \approx 220$ km s$^{-1}$, $M_{star} \sim 10^{11}$ $M_{\odot}$) less than 0.1% of blue galaxies are bulgeless, while for M33 mass galaxies ($V_{rot} \approx 120$ km s$^{-1}$, $M_{star} \sim 10^{10}$ $M_{\odot}$) bulgeless galaxies are more common, with 45% of blue galaxies having Sérsic index $n < 1.5$. Thus the challenge for ΛCDM is to produce enough M33-type galaxies [177].

## Small Scale Issues: Summary

**Satellites**: The discovery of many faint Local Group dwarf galaxies is consistent with ΛCDM predictions. Reionization, lensing, satellites, and Lyman-alpha forest data imply that if the dark matter is WDM, it must be tepid at most – i.e., not too warm.

**Cusps:** Recent high-resolution observations of nearby low-mass disk galaxies provide strong evidence that the central dark matter often has a nearly constant density core, not the NFW-type $\rho(r) \propto r^{-1}$ cusp. But the target is changing (which no doubt infuriates some observers), as high-resolution ΛCDM simulations including baryons appear to be producing dwarf spheroidal and low-mass spiral galaxies consistent with these observations. Better observations and simulations are needed.

**Disks:** ΛCDM simulations are increasingly able to form realistic spiral galaxies, as resolution improves and feedback is modeled more physically. However, accounting for the statistics on thin disks and bulgeless galaxies as a function of galaxy mass will be a challenge for continually improving simulations and semi-analytic models.

# 3. GALAXY DATA VS. SIMULATIONS

Structure forms by gravitational collapse in the expanding universe. If the universe were of uniform density, all gravity would do is to slow the expansion. Gravity needs initial inhomogeneities in order to generate structure. A power spectrum $P(k) = A\, k^n$ of roughly scale-invariant (i.e., with $n \approx 1$) Gaussian fluctuations is generated by quantum effects during inflation, and it can be arranged that the coefficient is of the required magnitude to generate the observed galaxies and large scale structure as these fluctuations grow and collapse. Although the average density of the universe falls as the universe expands, positive fluctuations expand slightly slower than average and grow steadily denser than average regions of the same size. After a given region reaches about twice the average density it stops expanding and collapses, as illustrated for an idealized spherical collapse in Fig. 8. Meanwhile the rest of the universe continues to expand around it. Through “violent relaxation” [178,179] the dark matter quickly reaches a stable configuration that is about half the maximum radius, with density falling with radius $r$ roughly as $r^{-2}$, corresponding to the observed flat rotation curves of spiral galaxies.

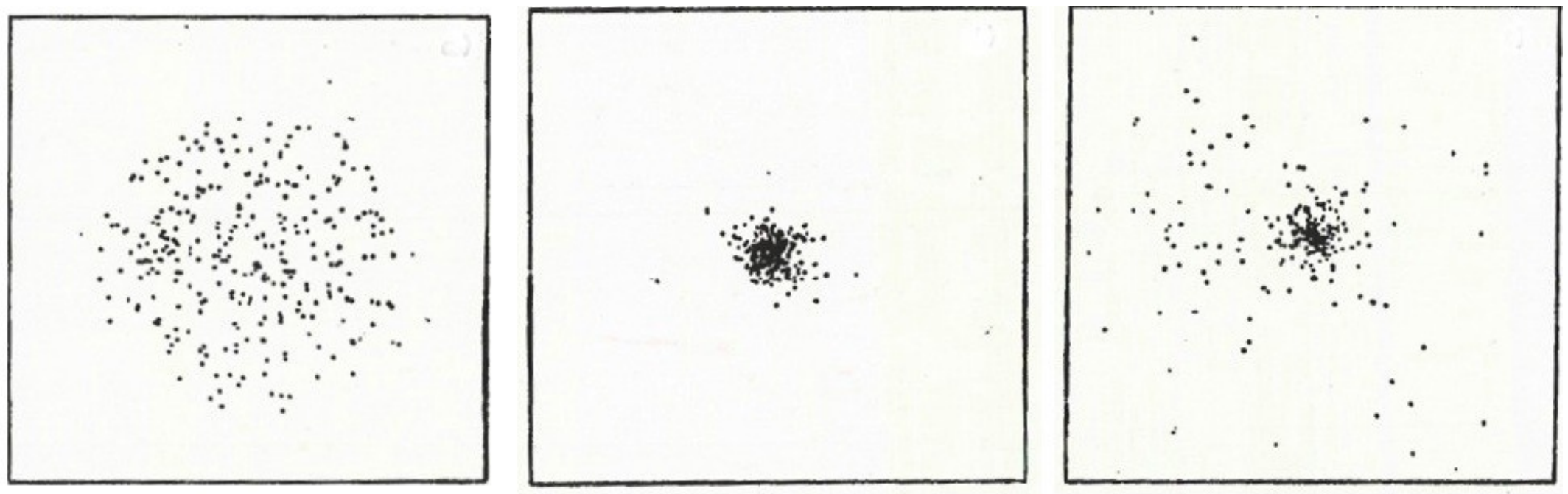

**FIGURE 8.** When any region becomes about twice as dense as typical regions its size, it reaches a maximum radius (**left panel**), stops expanding, and starts falling together. The forces between the subregions generate velocities that prevent the material from all falling toward the center (**center panel**). Through Violent Relaxation the dark matter quickly reaches a stable configuration that's about half the maximum radius but denser in the center (**right panel**). (From one of the first N-body simulations [180].)

Since fluctuations initially start with very small amplitude ($\sim 3\times10^{-5}$), their initial evolution can be followed by linear theory (Fig. 2), and there are now highly developed computer codes available to do this [181]. But once the amplitude approaches unity, nonlinear effects become increasingly important and we must resort to simulations.

Unlike the idealized example of an isolated halo in Fig. 8, a dark matter halo in CDM continues to accrete dark matter unless it is itself accreted by another dark matter halo. As mentioned in Lecture 2 in connection with the discussion of cusps at galaxy centers, the NFW [106] formula

$$\rho_{NFW}(r) = 4\ \rho_s\ x^{-1}\ (1 + x)^{-2}$$

is a rough fit to the dark matter radial density profile of simulated pure dark matter CDM halos. Here $x = r/r_s$, and the scale radius $r_s$ and the density $\rho_s$ at this radius are the two NFW parameters. The inner $r^{-1}$ part of the halo forms early, and then $r_s$ stays pretty constant as subsequently accreted dark matter is mostly kept away from the center by the angular momentum barrier. Thus the halo concentration $c_{vir} = R_{vir}/r_s$ grows with time; simulations show that $c_{vir}$ typically grows linearly with scale factor $a = (1 + z)^{-1}$ [150]. The average mass accretion history of halos is exponential in redshift $z$ [147], and the angular momentum parameter $\lambda$ of the halo typically grows significantly in halo major mergers (i.e., mergers with mass ratios between unity and ~1/3) and declines as mass is accreted in minor mergers [156]. Dark matter halos are generally triaxial spheroids; they are more elongated at smaller radii, larger redshifts, and higher masses [48], perhaps reflecting early accretion from narrow filaments, with accretion becoming more spherical as the filaments grow thicker than the halos. The Milky Way halo appears to be consistent with this [182]. (For recent more detailed studies and analytic approximations for halo properties and accretion histories, see [183-186] and references therein, and the forthcoming textbook by Mo, van den Bosch, and White [187].)

The baryonic component ($\Omega_b = 0.046$, with the universal baryonic fraction $f_b = 0.16$ of the average cosmic matter density $\Omega_m = 0.279$) can continue to radiate energy and fall toward the halo center, with a small fraction (usually less than 20%) of the baryons forming stars and becoming a visible galaxy. If the angular momentum distribution of the baryons were like that of the halo dark matter, the baryons would have a large central density peak and a very extended disk, and look nothing like observed baryons in galaxies [158,159]. But the baryonic angular momentum distribution need not be like that of the dissipationless dark matter since the baryonic matter behaves hydrodynamically: dark matter clumps interact only gravitationally and interpenetrate when they encounter each other, but baryonic clumps shock. As mentioned in Lecture 2, recent high-resolution hydrodynamical simulations [160,161] are starting to produce disk galaxies that are consistent with observations.

The distribution of galaxies is thus determined by the distribution of galaxy-mass dark matter halos, taking into account relevant astrophysical processes including merging of halos, gas heating and cooling, and star formation. The steadily increasing power of N-body simulations is shown by Fig. 9 (left), and the excellent agreement between the Millennium Run ΛCDM simulation and the observed galaxy 2-point correlation function is demonstrated in Fig. 9 (right). Note that the simulated and observed galaxies have significantly lower correlations on scales $< 2h^{-1}$ Mpc than dark matter particles. This is because of destruction of halos by tidal effects and interactions in dense environments, where most of the small-scale pairs of galaxies, halos, and particles are found. My colleagues and I had realized some time ago [188] that such "scale-dependent anti-biasing" must occur for ΛCDM to agree with observations, and it was indeed confirmed that this does occur when simulations of sufficient resolution became available [189-191].

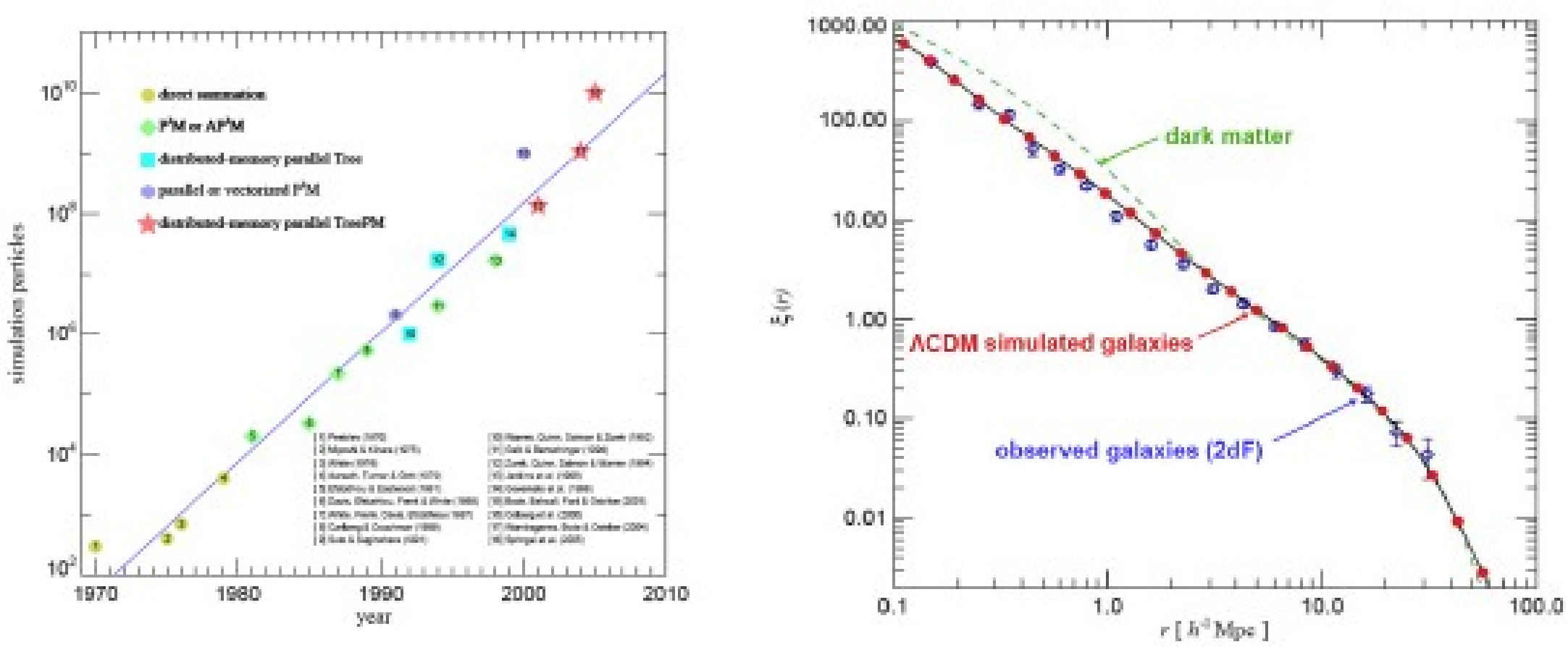

**FIGURE 9.** **(left)** N-body simulation particle number vs. publication date, showing exponential growth. **(right)** Galaxy 2-point correlation function at the present epoch, comparing observed galaxies from the 2dF redshift survey with simulated galaxies from the Millennium Run [191].

The Halo Model [192] is a simplified treatment of the evolution of large-scale density as a result of nonlinear gravitational clustering. The Halo Occupation Distribution (HOD) formalism [193] is based on the assumptions that all galaxies occupy dark matter halos, and that the number of galaxies within a halo brighter than a given luminosity depends only on the mass of the halo and not (to a first approximation) on its larger-scale environment.

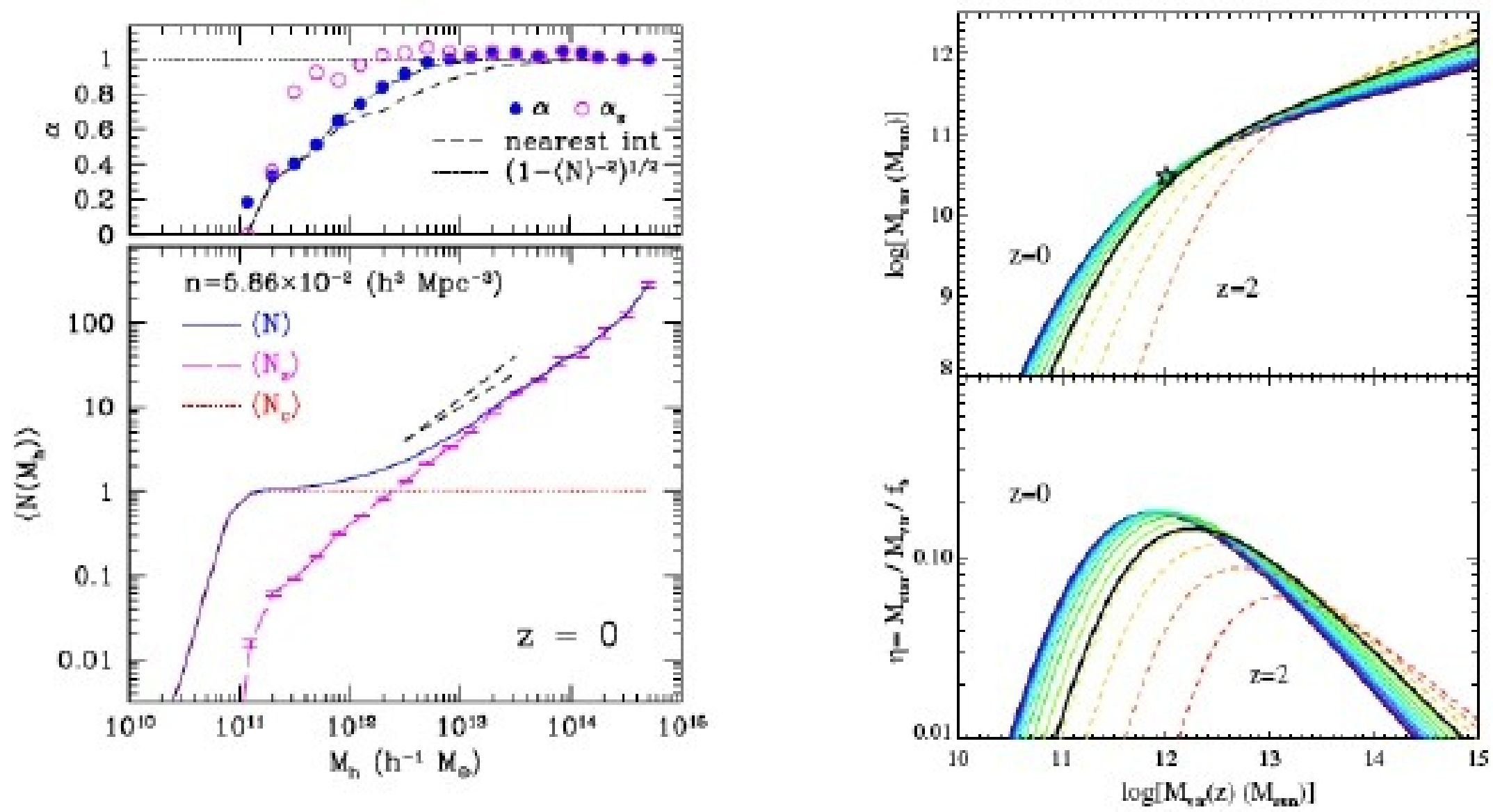

**FIGURE 10**. (**left**) Halo occupation of galaxy subhalos in their hosts. (**lower left panel**) First moment of the halo occupation distribution of subhalos, as a function of host mass, at $z = 0$. The plot shows the mean total number of halos including the hosts (solid line), the mean number of satellite halos (long-dashed line), and the step function corresponding to the mean number of "central" halos (dotted line). The two short-dashed lines indicate scaling with $M_h$ and $M_h^{0.8}$. (**upper left panel**) The parameter $\alpha \equiv \langle N(N-1)\rangle^{1/2}/N$ for the full HOD (solid points) and the HOD of satellite halos (open points). The dotted line at $\alpha = 1$ shows the case of a Poisson distribution. The full HOD at small $M_h$ is described by the nearest integer distribution (dashed line). (Figure from [190].) (**upper right panel**) The relation between galaxy stellar mass and halo mass from $z = 2$ to $z = 0$, using the abundance matching model. (**lower right panel**) Fraction of available baryons that exist in stars, as a function of the halo mass and redshift, where $f_b$ is the universal baryon fraction. The star marks the location of the Milky Way at $z = 0$. The thick black line represents the relation at $z = 1$. (Figure from [192].)

It is reasonable to assume that the brightest galaxies in a halo occupy the subhalos with the largest maximum circular velocity $V_{max}$. High-resolution ΛCDM cosmological simulations were analyzed using the HOD formalism with this assumption [190], as illustrated in Fig. 10 (left). We found that the number of galaxies brighter than a given luminosity scales with the halo mass $M_h$, plus a central galaxy. This analysis led to the prediction that the short-range autocorrelation function of halos that host galaxies becomes steeper at higher redshifts, as illustrated in Fig. 11.

Such predictions appear to be in excellent agreement with observations [191] when galaxies are associated with halos according to a simple prescription [192,193] in which galaxies ranked by luminosity are matched to dark matter halos or subhalos ranked by $V_{max}$ (for subhalos, $V_{max}$ at the time of accretion). This is true both for relatively nearby galaxies in the Sloan Digital Sky Survey (SDSS), and also for galaxies at redshifts $z \sim 1$ and even $z \sim 4$, as shown in Fig. 12. The same simple abundance matching model can reproduce other galaxy clustering statistics, including close-pair evolution [194], galaxy-mass correlations from weak gravitational lensing [195], and three-point correlations [196].

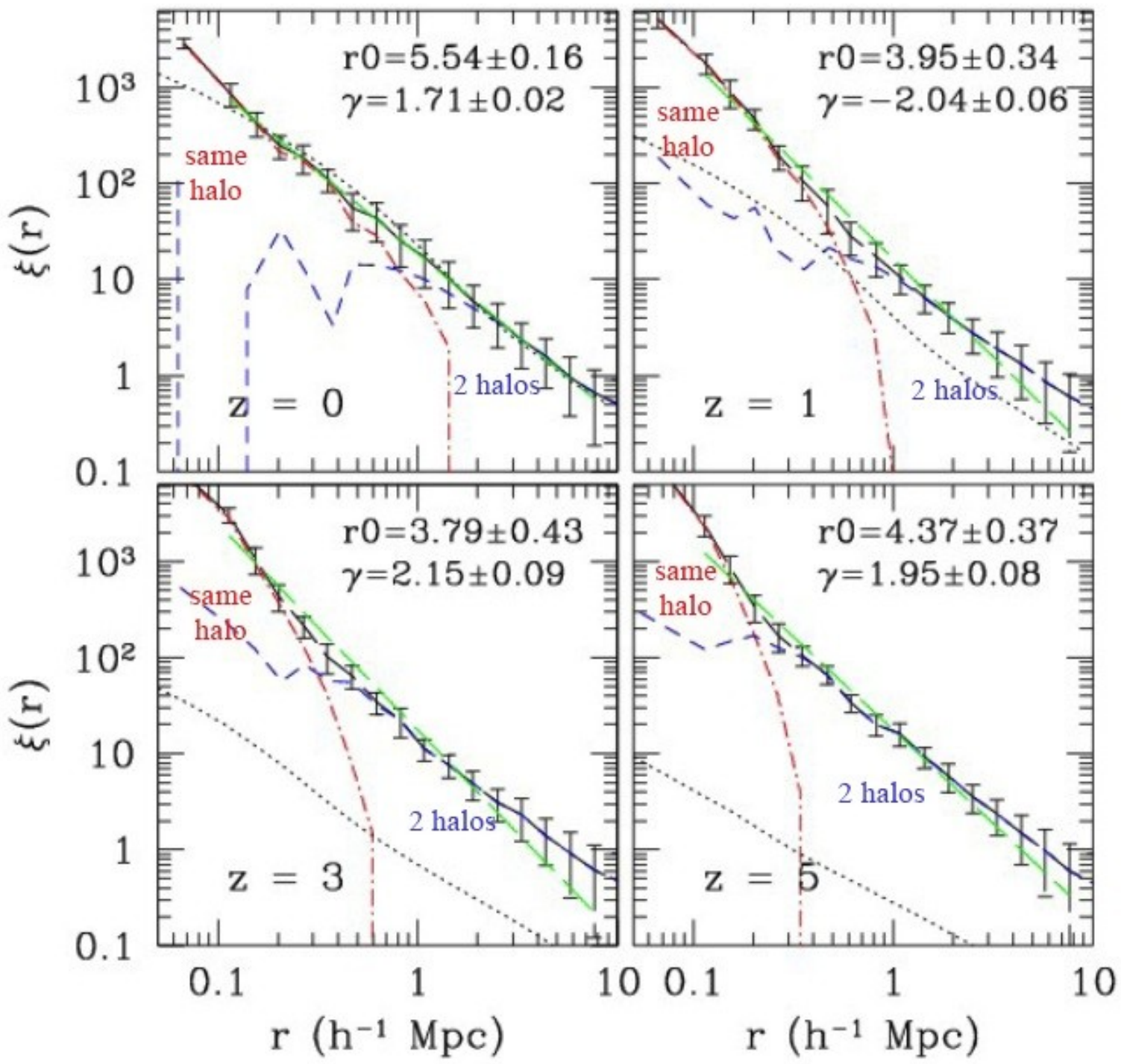

**FIGURE 11.** Evolution of the 2-point correlation function. The solid line with error bars shows the clustering of halos with fixed number density $n = 5.89\times10^{-3}\ h^3\ \mathrm{Mpc}^{-3}$ at redshifts $z = 0$, 1, 3, and 5. The dot-dashed and dashed lines show the corresponding one- and two-halo term contributions in the HOD analysis; the dotted line is the dark matter correlation function. The long dashed lines are power-law fits to the correlation functions in the range from 0.1 to 8 $h^{-1}$ Mpc. The correlation function steepens significantly at small scales $r < 0.3\ h^{-1}$ Mpc due to the one-halo term. (From [189].)

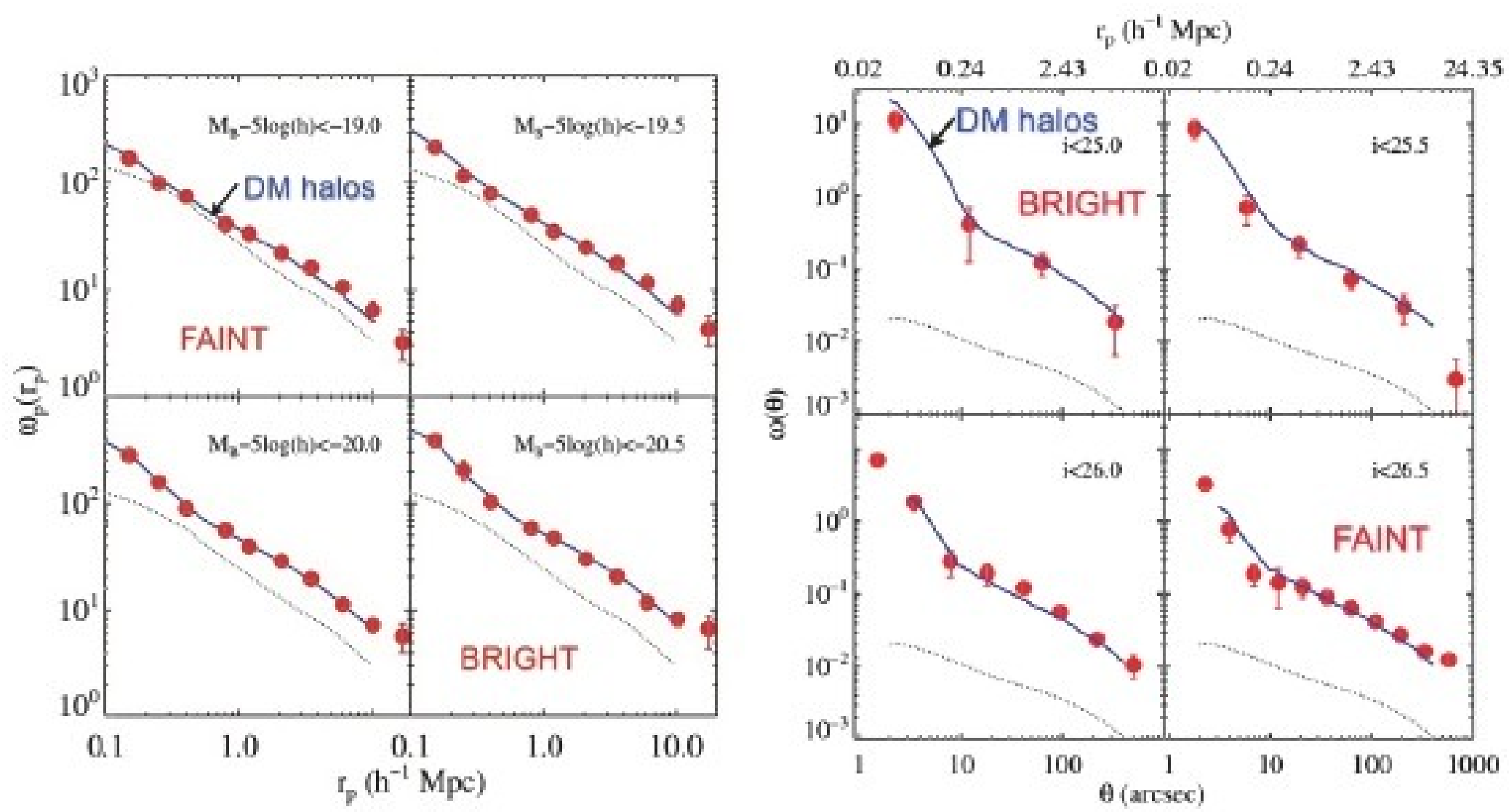

**FIGURE 12.** Observed galaxy clustering agrees with ΛCDM simulations at redshifts $z \sim 1$ (**left**) and at $z \sim 4$ (**right**). The circles are projected correlation functions from DEEP2 observations at z ~ 1 and Subaru observations at z ~ 4, while the curves are the correlation functions of halos at the same number density. (From [191].)

The success of the abundance matching model has allowed Conroy and Wechsler [192] to attempt to understand the evolution of the buildup of stellar mass and the implied star formation rate of galaxies as a function of their halo mass from the nearby universe out perhaps to $z \sim 2$. The mass of a subhalo is taken to be its mass at the time of accretion. Results are shown in Fig. 10 (right), in which the redshift curves are spaced equally in scale factor $a = (1 + z)^{-1}$. In the upper panel, the cross-over of these curves at halo mass $M_{\rm vir} \approx 10^{12.5}$ $M_\odot$ implies that halos of this mass host nearly constant stellar mass $M_{\rm star} \sim 10^{11}$ $M_\odot$, while halos grow faster than their stellar contents above this halo mass and slower below it. The Milky Way (star symbol) lies on the mean relation. The lower right panel shows the star formation efficiency η, defined as the ratio of stellar mass to the total baryonic content of the halo ($f_b M_{\rm vir}$). The overall efficiency of converting baryons to stars is quite low, with a peak efficiency of ~20% at $z \sim 0$ at $M_{\rm vir} \sim 10^{12}$ $M_\odot$. This low efficiency is in good agreement with estimates for the Milky Way [197], with estimates based on weak lensing [198], and with accounting of baryons in various states [42,199].

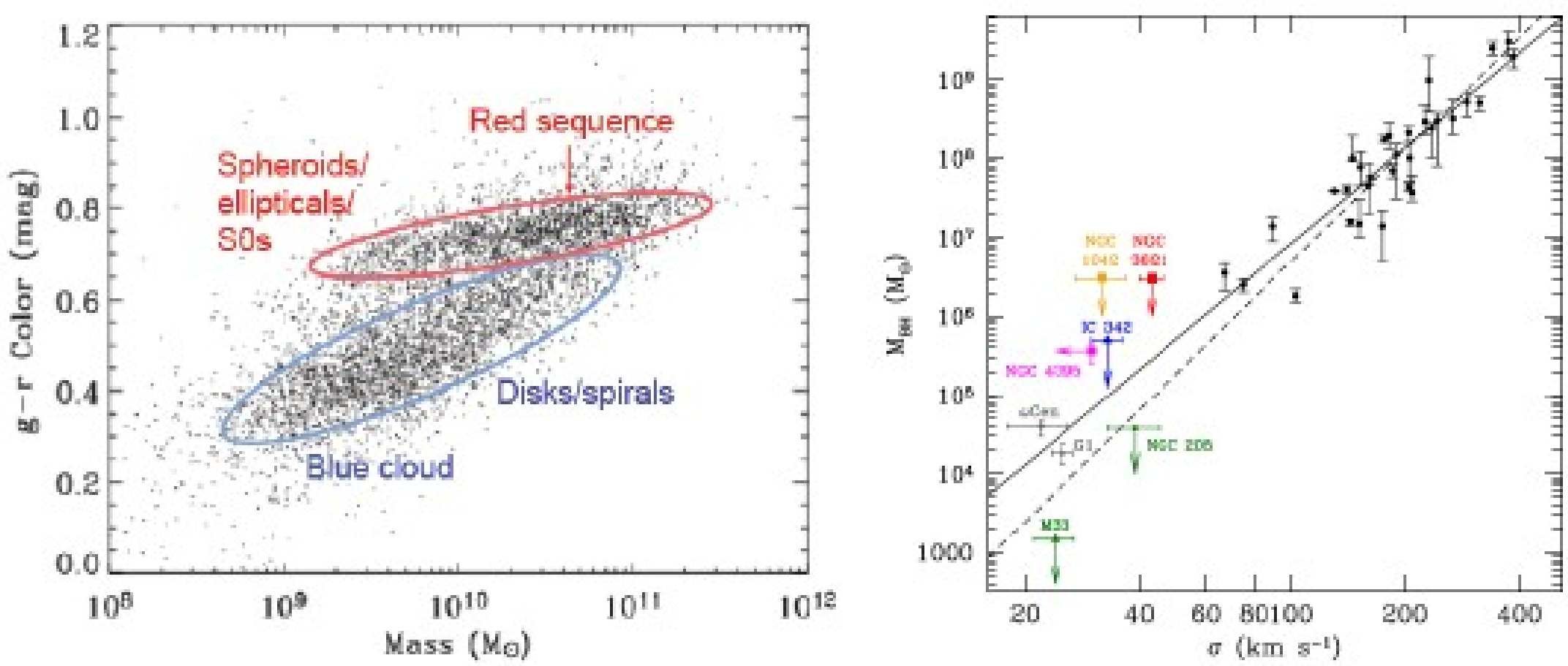

**FIGURE 13.** (**left**) Color vs. stellar mass for SDSS galaxies, showing color bimodality. (**right**) The $M_{\rm bh}$-$\sigma_*$ relation including nuclear clusters in late-type spiral galaxies (from [214]). Black circles are from [212]. Filled squares represent nuclear clusters containing AGNs, and triangles represent nuclear clusters in late-type spirals without AGNs.

A key feature of the galaxy population is the dichotomy between blue, often disky, star-forming galaxies and red galaxies with an older, often spheroidal, stellar population, as shown in Fig. 13 (left). This basic fact about nearby galaxies became especially clear through analyses of SDSS data (e.g. [200-202]). More distant surveys showed that the co-moving number density of blue galaxies has remained roughly constant since $z \sim 1$, while the number density of red galaxies has been rising [203,204]. UV-based star formation rate (SFR) measurements from the GALEX satellite agree well with SDSS Hα SFR measurements for nearby galaxies, and there appears to be an evolutionary sequence for massive galaxies that connects normal star-forming galaxies to quiescent galaxies via strong and weak AGN [205]. There is a "main sequence" of star forming galaxies out at least to $z \sim 1$ [206], with a relatively narrow range of SFRs at given stellar mass and redshift (the 1σ range is only about 0.3 dex), and with the average SFR increasing at higher redshift. The basic pattern seems to be that massive galaxies form stars early and fast, and are red today, while lower mass galaxies form stars later and more slowly: "staged" galaxy formation [207].

Although these data show that galaxy mergers cannot boost star formation very much at these redshifts, it has long been known from simulations (e.g. [208,209]) that major mergers can transform stellar disks into stellar spheroids. It is also now established that stellar spheroids host massive black holes, with the black hole mass $M_{\rm bh}$ approximately three orders of magnitude less than the spheroid mass [210] and $M_{\rm bh}$ scaling with the 4$^{\rm th}$ power of the central stellar velocity dispersion $\sigma_*$ [211-213]. Whatever transforms galactic stellar disks into spheroids therefore must also grow massive black holes, resulting in the release of large amounts of radiation. Such AGN activity may also heat the remaining galactic gas and remove some of it, thus quenching star formation and turning the galaxy red. But how this happens and what keeps the galaxy from subsequently forming new stars remain mysterious. Fig. 13 (right) shows that the $M_{\rm bh}$-$\sigma_*$ scaling may continue down to much lower $M_{\rm bh}$ with nuclear star clusters.

A large program of galaxy merger simulations by my group [215-218] and that of Lars Hernquist [219-220], many of them run by my former student T. J. Cox, has clarified the morphological transformations of galaxies during mergers and the possible role of mergers in producing bright AGN (quasars) and massive black holes, reviewed in [221,222] and illustrated in Fig. 14. Processing merger simulations with the Sunrise radiative transfer code [224-226], we are now determining the time scales [227-229] over which merger stages will be visible via close pairs [230] and using asymmetry [231] and Gini-$M_{20}$ [232] to measure galaxy morphology, and comparing to observations [233] to measure galaxy merger rates [234].

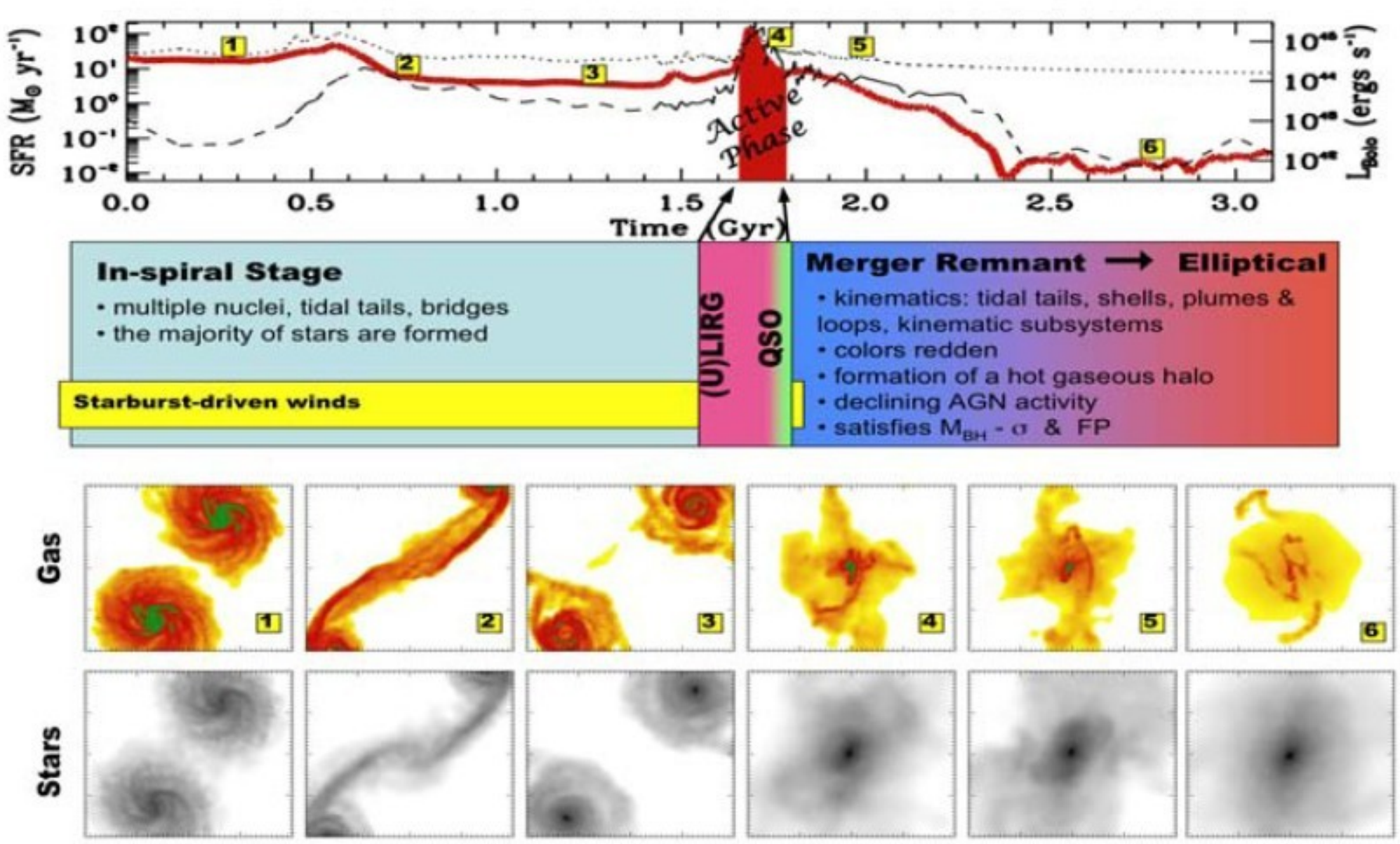

**FIGURE 14.** Schematic chronology of a gas-rich major merger: in-spiral stage, ultra-luminous infrared galaxy (ULIRG) followed by a brief QSO stage, leaving a quenched elliptical galaxy as the merger remnant. The top panel shows the star formation (left axis, thick solid line) and luminosity evolution (dashed line for the black hole, dotted line for the stars). Images of gas and stars at the numbered stages illustrated in the top panel are in the bottom panels. (From [223].)

# 4. SEMI-ANALYTIC MODELS OF GALAXY FORMATION

The original CDM paper [18] used a spherical gravitational collapse model of galaxy and cluster formation and a simplified theory of gas cooling in order to allow comparison with observations. But the key paper [168] that initiated modern semi-analytic models (SAMs) of galaxy formation was based on the extended Press-Schechter [235] theory [236-240] of dark matter halo merging and a more elaborate model of gas cooling by radiation and gas heating by gravitational collapse and stellar feedback. This was the basis for the first SAM papers [241,242], which assumed that most star formation occurs in galactic disks, that galactic stellar spheroids form only in major mergers, and that gas cools only onto the central galaxy in any halo, and used local data to adjust the parameters that describe star formation and feedback. These models reproduced remarkably well the observed trends in galaxy luminosity, gas content, and morphology, including that early-type galaxies (dominated by stellar spheroids) populate higher density environments in agreement with the observed density-morphology relation [243,244]. Kauffmann et al. [241] also pointed out that many low-mass dark matter halos must be underluminous in order not to produce more stellar light than is observed, an early prediction of the small number of luminous satellite galaxies compared to satellite halos that I discussed in Lecture 2.

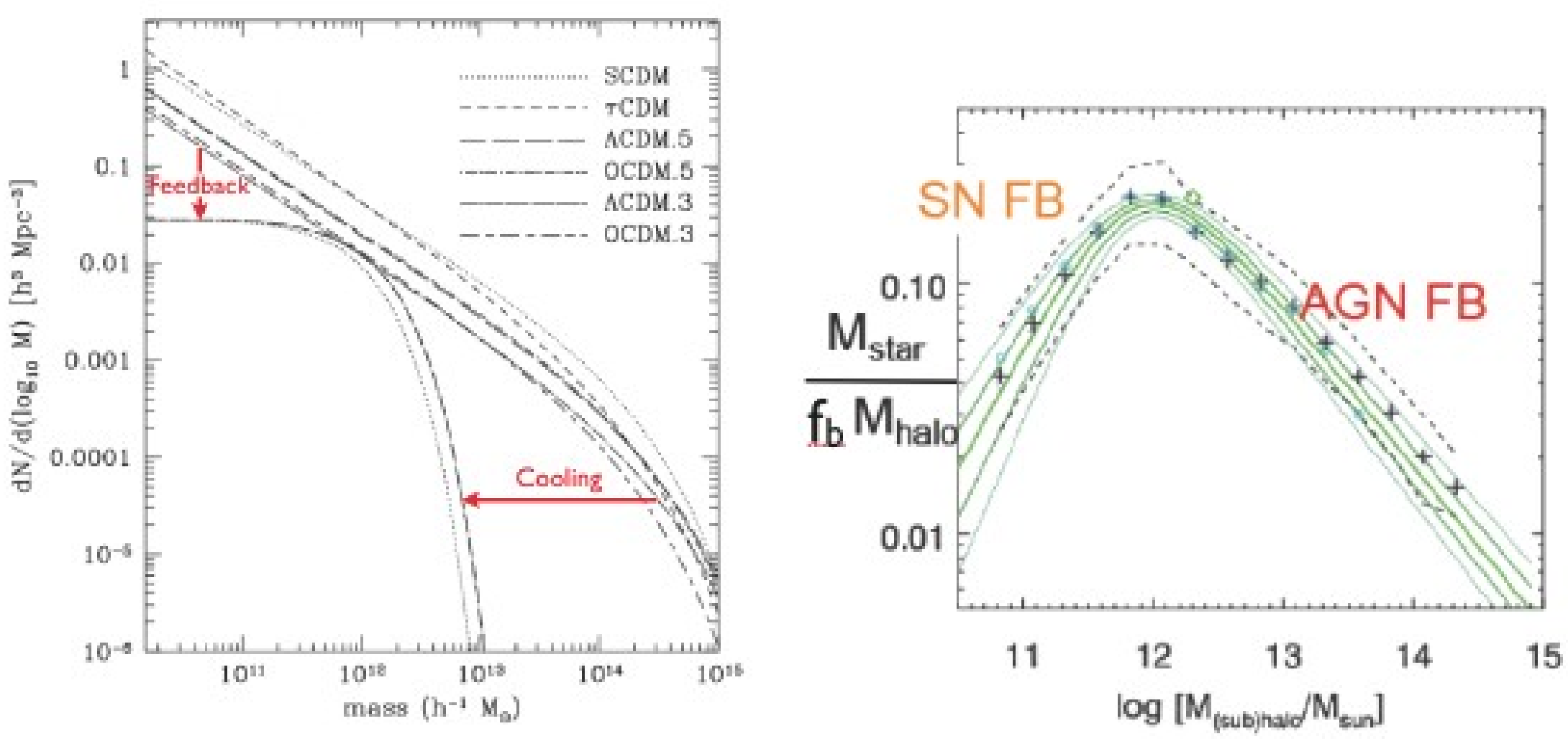

**FIGURE 15.** (**left**) All halos vs. galactic halos. The curves at the right show the Press-Schechter mass function of all halos for various cosmologies, while the curves below them show the mass function of halos hosting observed galaxies. (From [245].) (**right**) Stellar fraction of halo baryons as a function of halo or subhalo mass. The solid green lines show the empirical relation (with 1- and 2-σ errors), and + symbols show predictions of Somerville's recent SAM [246], which includes feedback from both supernovae and AGN. The dashed lines show the $16^{th}$ and $84^{th}$ percentiles for the fiducial model. (This is a simplified version of Fig. 3 of [246].)

That there are far more dark matter halos than halos hosting galaxies is evident from Figure 15 (left). (The numbers and masses of halos hosting galaxies could be estimated, for example, from galaxy luminosity functions plus the empirical Tully-Fisher and Faber-Jackson relations [247].) Indeed, it was anticipated from the beginning of CDM modeling that galaxy formation would be efficient only for dark matter halos in the mass range roughly $10^8$ – $10^{12}$ $M_{\odot}$, which lie below the cooling curves in Fig. 3. Including dust extinction [245,248] helped SAMs to reproduce observed luminosity functions. However, SAMs typically overproduced very luminous galaxies unless additional astrophysics was invoked, such as AGN feedback (e.g. [249]). Croton et al. [250] additionally added "radio-mode" AGN feedback to their SAM based on the Millennium Run simulation in order to quench star formation by keeping hot gas from cooling, and succeeded in predicting galaxy color bimodality with the most massive galaxies at $z \sim 0$ red, in agreement with observations. An alternative scenario for quenching of star formation implemented in another SAM [251,252] appeals to the existence of a critical halo mass $M_{\mathrm{h,crit}} \sim 10^{12}$ $M_{\odot}$ such that gas can enter halos with $M_{\mathrm{h}} < M_{\mathrm{shock}}$ in cold streams and form stars efficiently, while gas entering halos more massive than $M_{\mathrm{shock}}$ at $z < 2$ is shock-heated and cannot form stars efficiently [165-167]. Figure 15 (right) shows that a modern SAM [249] including supernova feedback (SN FB) plus AGN feedback using prescriptions based on simulations [221,222] is in good agreement with the observed star-formation efficiency as a function of halo mass, already discussed in connection with the lower right panel of Fig. 10. The left panel of Fig. 16 shows the distribution of dark matter halos in the vicinity of a rich cluster at $z = 0$ from the Millennium Run, and the right panel shows that the central galaxies are all red, consistent with observations.

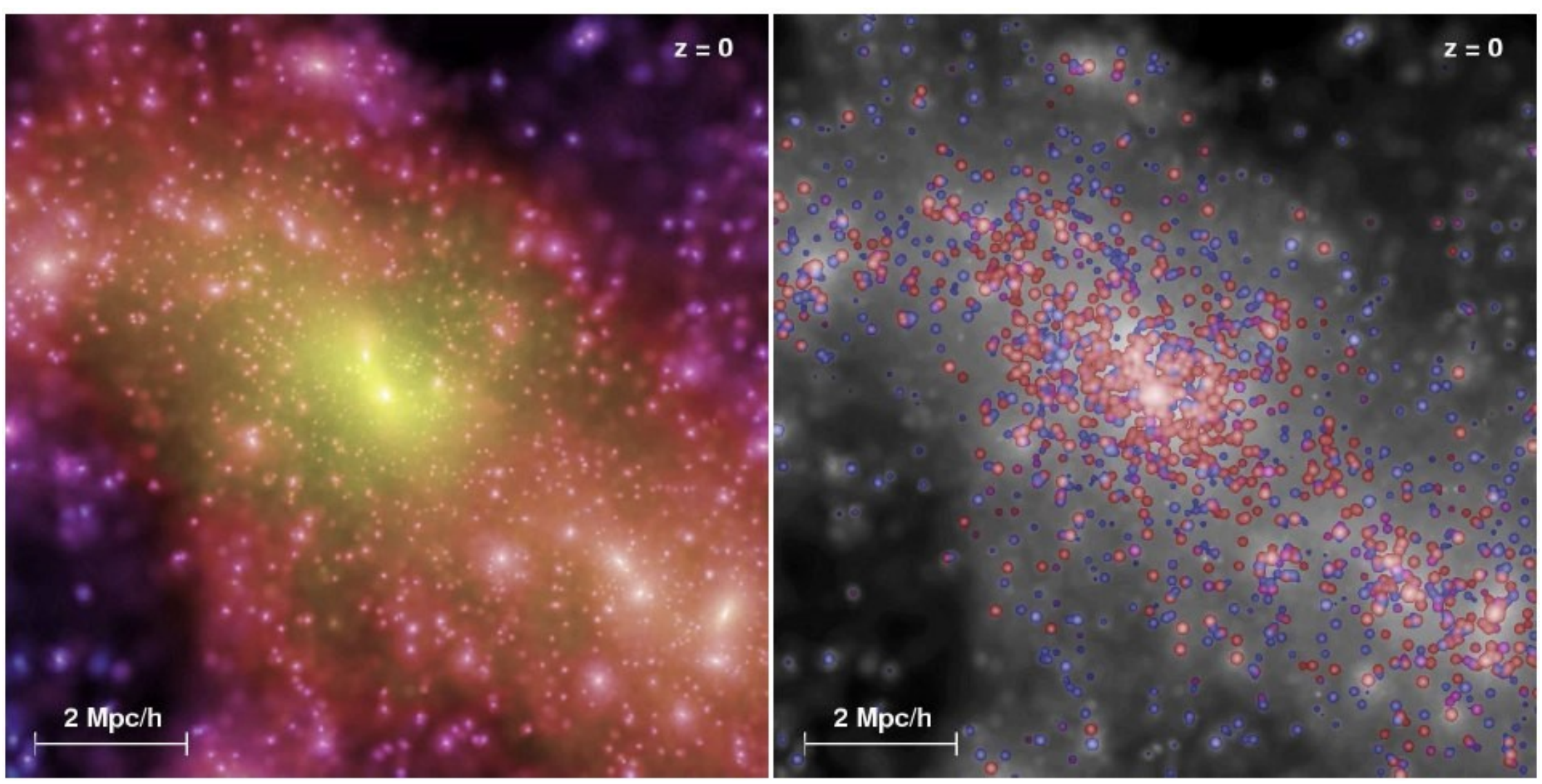

**FIGURE 16.** Projected dark matter and galaxy distribution centered on a rich cluster from the Millennium Run. (**left**) Projected dark matter distribution in a 10 $h^{-1}$ Mpc cube, with color representing velocity dispersion and brightness representing dark matter density. (**right**) Same volume, now showing dark matter density in grayscale and galaxies in the SAM with the colors representing stellar restframe color and the sphere volume proportional to the galaxy's stellar mass. (From [191].)

As mentioned at the end of Lecture 3, major mergers can transform disks into spheroids and lead to bursts of star formation and rapid growth of black holes when sufficient gas is available. But it is important to appreciate that most star formation does *not* occur in starbursts. This is shown in Fig. 17 (left), where the upper dashed and dot-dashed curves show the total predicted star formation rate density, and the lower fainter curves show the SFR due to bursts.

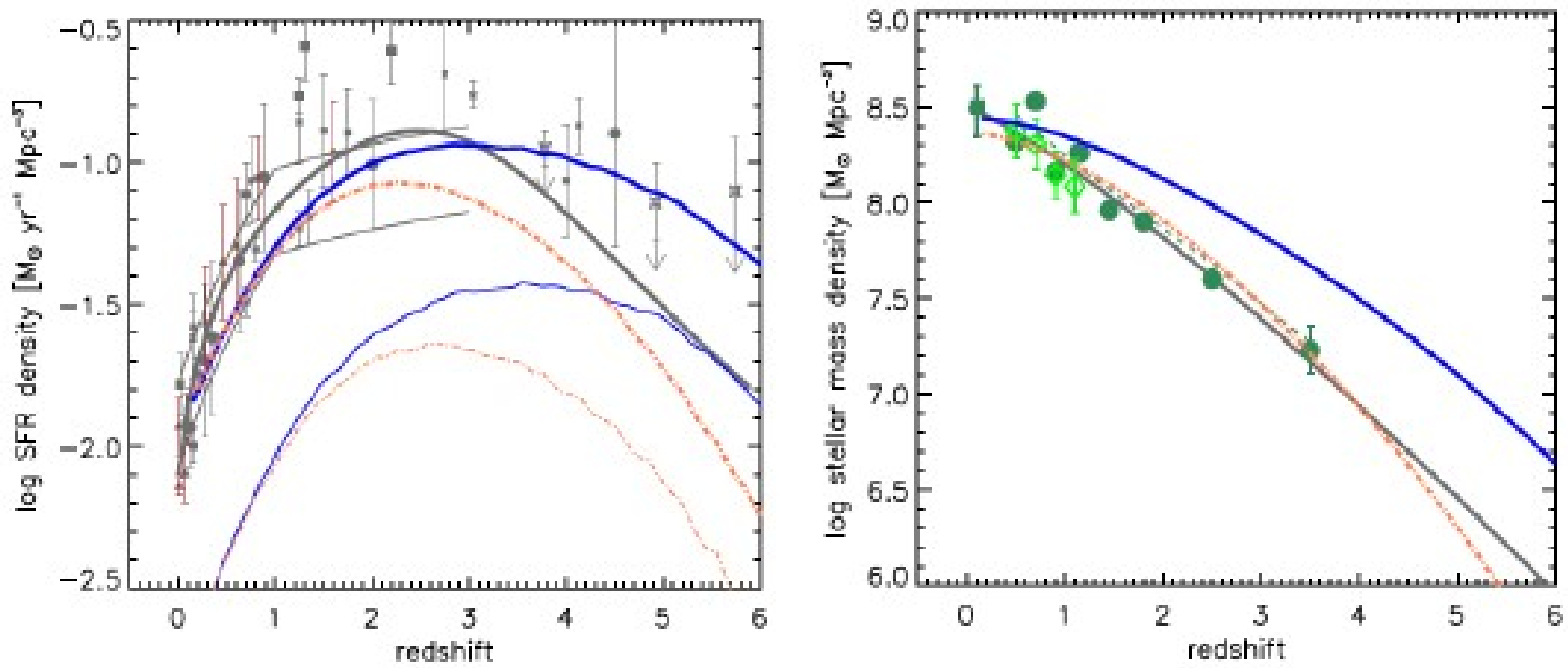

**FIGURE 17.** (**left**) Star formation rate density as a function of redshift (Madau plot). The upper solid blue curve shows the total formation according to the 'fiducial' SAM of [246], while the lower light blue curve shows the SFR in bursts. The dot-dashed orange curves show the total and bursty star formation in a 'low' model with reduced star formation in small-mass halos. The symbols and solid lines show observational results converted to a Chabrier IMF (see [253,254] and [246] for details). (**right**) The integrated global stellar mass density as a function of redshift (Dickinson plot). Symbols and the solid grey line show observational estimates (see [246]). The upper blue curve shows the 'fiducial' model, and dot-dashed red curve shows the 'low' model. (From [255].)

The predicted integrated stellar mass density from the same models is shown in Fig. 17 (right). Although the 'fiducial' model (dashed curve) is a better fit to the SFR density, the 'low' model (dot-dashed curve) is a better fit to the stellar mass density. This shows that there is an inconsistency between these two quantities, despite the fact that the SFR density integrated over time should equal the stellar mass density. Clearly, it is difficult to assess the success of models of the evolving galaxy population unless this discrepancy can be resolved. One possible resolution could be that the stellar initial mass function (IMF) evolves, becoming more top-heavy (i.e., producing a higher fraction of high-mass stars) with increasing redshift [254,256, 257]. Another possibility is that the SFR in Fig. 17 (left) was overestimated at higher redshifts (e.g. [258]).

Modern SAMs reproduce many features of the observed universe very well, showing that they are getting some aspects right. But it is always important to ask what they get wrong, since that may lead to progress. One persistent problem is getting star formation right in small galaxies, such that lower-mass galaxies have higher specific star formation rate (SSFR = SFR per unit stellar mass), as observed (e.g. [246]). Another problem is getting the black hole accretion history right. Both simulations and SAMs can correctly reproduce the observed correlation of supermassive black hole mass with stellar spheroid mass. But although Hopkins et al. [221] claimed good agreement with the observed evolution of quasar luminosity density, SAMs [246,250] do not reproduce this. The difference turned out to be that the simpler calculations summarized in [221] allowed the black holes to grow from small seed masses to the final mass in each merger, while the SAM [246] based on the galaxy merger simulations [221] treated the black holes self-consistently, starting from the black holes grown previously [259]. A recent review [260] summarizes much data and many questions that are still open concerning the evolution and role of supermassive black holes in galaxies and clusters.

It is worth emphasizing how well a simplified SAM does that is based on the idea that star formation is only efficient for ΛCDM halos in a narrow mass range from $M_{min}$ to $M_{max} = M_{shock} = 1.5\times10^{12}$ $M_{\odot}$ [261]. The specific star formation rate is assumed to be approximately equal to the universal baryon fraction $f_b$ times the mass accretion rate of halos [262-264], which is shown to be consistent with observations in Fig. 18 (left). The resulting Madau plot is shown in Fig. 18 (right), and one can see that $M_{min} = 10^{11}$ $M_{\odot}$ matches the observations well.

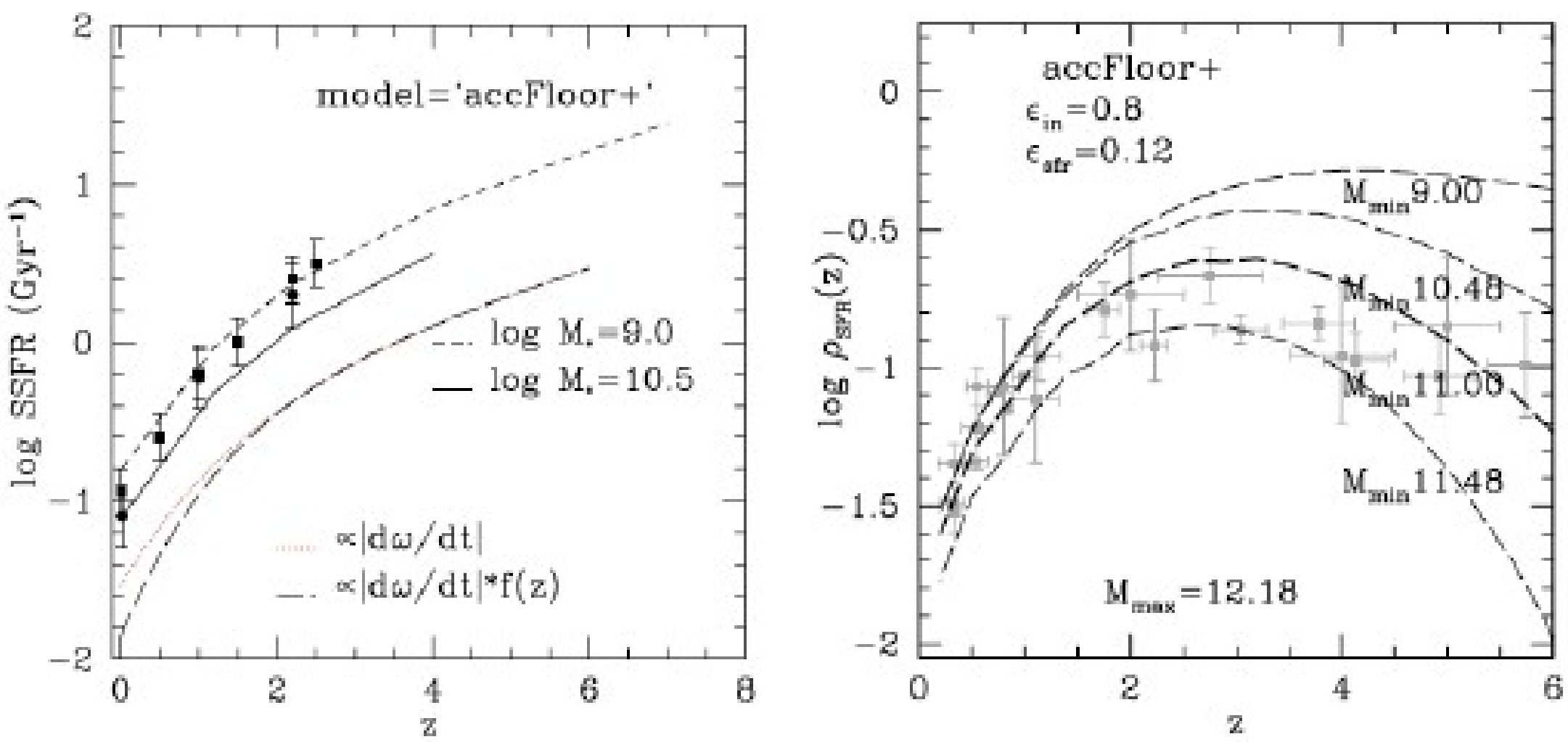

**FIGURE 18.** (**left**) Specific star formation rate as a function of redshift for stellar mass $M_* = 10^9$ and $10^{10.5}$ $M_{\odot}$ (short-dashed and solid curves) from the fiducial model of [261] compared to compilations of recent observations for $M_* = 10^{10.5}$ $M_{\odot}$ (solid circles [265], solid squares [266]). The star formation rate is assumed to be proportional to the dark matter halo growth rate known from simulations [262,263] and the EPS formalism [264], which in turn is governed by $\omega(z) = 1.69/D(z)$ where $D(z)$ is the linear growth factor. (**right**) Star formation rate vs. redshift (Madau plot) from the fiducial model of [260] (dashed curves) compared with recent observations. A Kennicutt model corresponding to star formation efficiency $\varepsilon_{sfr} = 0.12$ is used. (Figures from [261].)

Cowie et al. [267] defined "downsizing" as "the remarkably smooth downward evolution in the maximum luminosity of rapidly star-forming galaxies," resulting in the assembly of the upper end of the galaxy luminosity function occurring from the top down with decreasing redshift. That massive galaxies form their stars first initially seemed at odds with the hierarchical nature of the cold dark matter paradigm, in which small halos form first and agglomerate into larger ones. But the idea that star formation is efficient only in dark matter halos with a narrow range of masses naturally explains how the phenomenon of downsizing arises: halos that are massive today passed through the star forming mass band between $M_{min}$ and $M_{shock}$ earlier and thus formed their stars earlier than halos that are less massive today. My UCSC colleague Sandra Faber likes to make an analogy between galaxies and stars: mass is destiny for both.

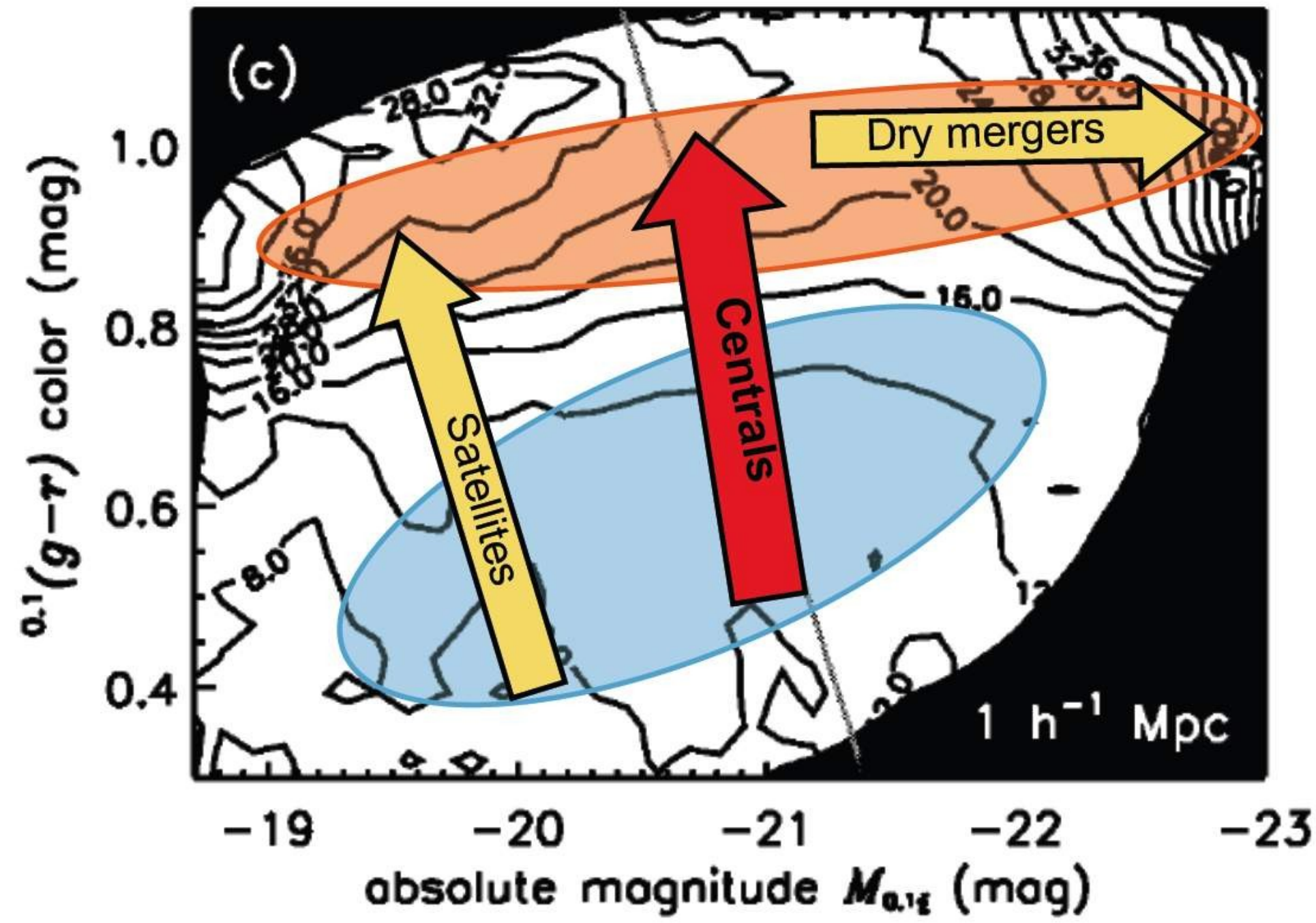

**FIGURE 19.** Color-magnitude diagram of nearby galaxies from the SDSS, showing contours of mean overdensity in spheres of 1 $h^{-1}$ Mpc. (From [268], with blue cloud, red sequence, and arrows added by S. M. Faber.)

Figure 19 illustrates schematically how galaxies evolve. Galaxies forming stars are in the blue cloud. Some galaxies have their star formation quenched when they become satellite galaxies in a larger halo, they cease to accrete gas, and they join the red sequence. Central galaxies form in the blue cloud, but they join the red sequence when they form a supermassive black hole and/or their halo mass exceeds approximately $M_{\rm shock}$ and/or they become satellite galaxies in a cluster. The most massive red galaxies cannot have simply be quenched central blue galaxies, since the latter are not massive enough; thus they must have been created by mergers without much star formation, which Rachel Somerville calls "dry mergers."

Modern SAMs can rather accurately reproduce the observed galaxy luminosity functions out to high redshift, and they capture at least a significant fraction of the relevant astrophysical processes. Therefore, with an adequate treatment of absorption and reemission of light by dust, such models can be used to calculate the extragalactic background light (EBL). This is important, since the burgeoning field of gamma-ray astronomy is providing increasingly restrictive upper limits on the EBL from the optical to mid-infrared wavelengths. The connection between the EBL and gamma rays arises because the main physical mechanism that attenuates high-energy gamma rays on their way from sources such as blazars (AGNs with relativistic jets pointing at the observer) or gamma ray bursts (GRBs) to our telescopes is pair production: $\gamma\gamma \rightarrow e^+e^-$. The gamma ray energy tunes the EBL wavelength range probed; for example, when a 1 TeV gamma ray hits a 1 eV photon of starlight (with wavelength ~1 μm) the center-of-mass energy is 1 MeV, enough to create an $e^+e^-$ pair.

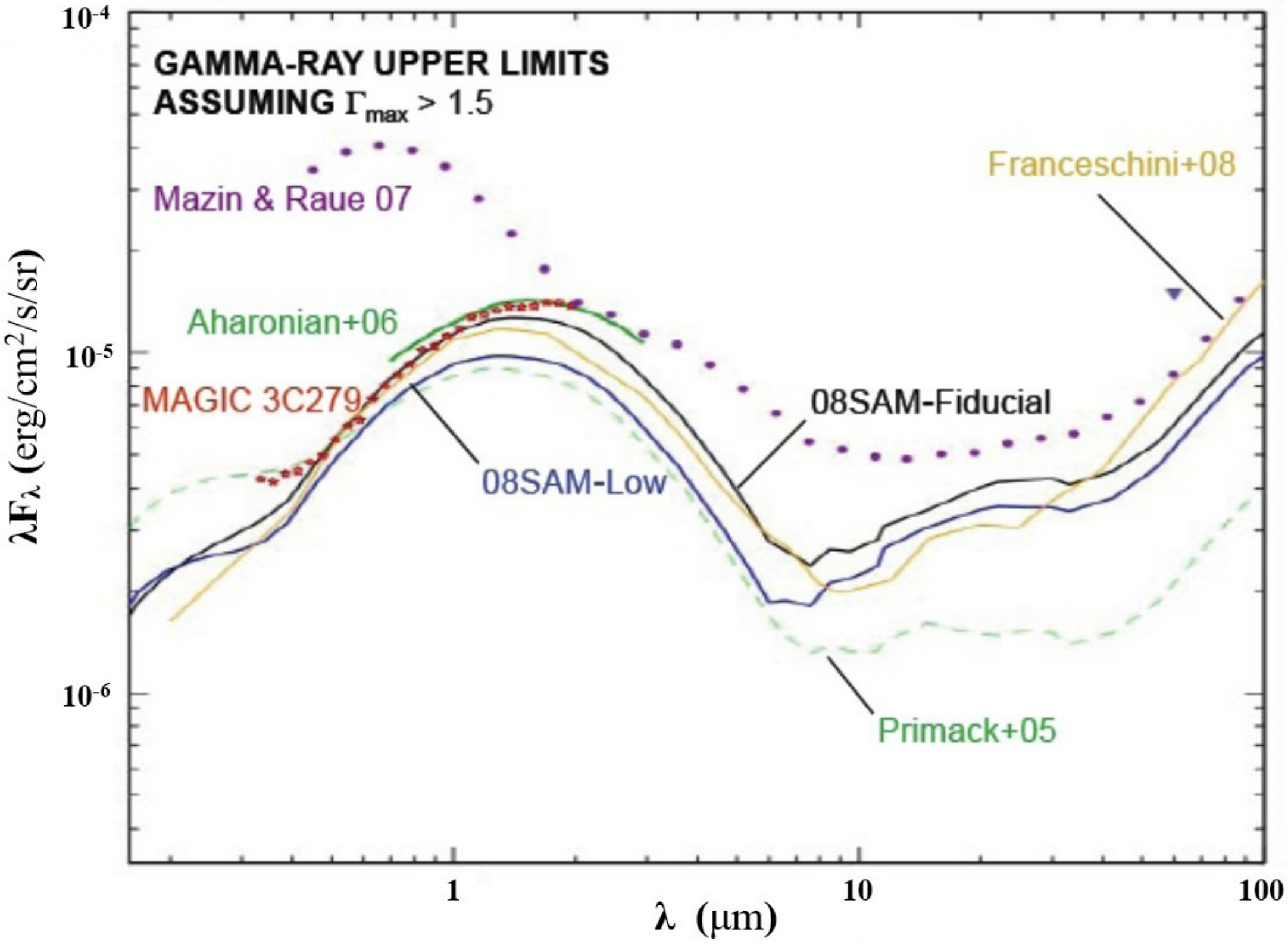

**FIGURE 20**. Extragalactic background light (EBL) from the 'fiducial' and 'low' models [246] illustrated in Fig. 17, including upper limits on the EBL from several blazars at $z \sim 0.2$ and a quasar at z = 0.53. The curve labeled Primack+05 is from [269], which cautioned that its treatment of dust emission was inadequate for wavelengths longer than ~10 μm, and the curve labeled Franceschini+08 is and observationally based backward evolution estimate of the EBL from [270]. The dotted and green upper limits on the EBL [271-273] are discussed in the text.

The 'fiducial' and 'low' Somerville et al. [246] SAMs discussed above in connection with the Madau and Dickinson plots in Fig. 17 lead to the EBL curves labeled 08SAM-Fiducial and 08SAM-Low in Fig. 20, which bracket theoretical expectations. (These results are similar to our new, improved EBL calculation [274, 275], including better modeling of ionizing radiation [276] and using the new *Spitzer* dust-emission templates [277].) Since gamma-ray attenuation increases with gamma-ray energy $E_\gamma$, upper limits to the EBL can be obtained by assuming that the un-attenuated gamma-ray energy spectra from sources including $z \sim 0.2$ blazars and QSO 3C279 at z = 0.53 are not harder than $E_\gamma^{-\Gamma}$ where $\Gamma > 1.5$. This is plausible since nearby, relatively un-attenuated sources typically have $\Gamma > 2$. Three limits from recent papers are plotted in Fig. 20. These show that there is little room in the optical and near-infrared energy range for additional sources of extragalactic background light beyond those included in the SAMs we have used, which potentially constrains the entire history of galaxy formation (e.g. [256]). With the *Swift* and *Fermi* gamma-ray satellites continuously monitoring the sky for GRBs and flaring blazars, and with ground-based atmospheric Cherenkov telescope arrays such as H.E.S.S., MAGIC, and VERITAS steadily gaining observing power, such constraints on the EBL can be expected to improve significantly in the near future.

## ACKNOWLEDGMENTS

I thank my collaborators – especially Avishai Dekel, Sandra Faber, and Rachel Somerville – for many helpful discussions, and for some of the slides used in the lectures that I gave in Rio. I also thank my current and former students and other colleagues, including the participants in the 2009 Caltech workshop supported by the W. M. Keck Institute for Space Studies (KISS) "Shedding Light on the Nature of Dark Matter" (http://www.kiss.caltech.edu/mini-study/darkmatter/index.html) and also the participants in the 2009 University of California Santa Cruz Workshop on Galaxy Formation (http://physics.ucsc.edu/SCGW09/), for many helpful discussions. I am grateful to NASA and NSF for grants that supported my research relevant to topics covered in these lectures.